\documentclass[aps,prb, reprint, superscriptaddress,secnumarabic,amssymb,nobibnotes]{revtex4-2}
\usepackage{graphicx}
\usepackage{amsmath}
\usepackage{subfigure}
\usepackage{verbatim}
\usepackage[english]{babel}
\usepackage[utf8]{inputenc}
\usepackage[colorinlistoftodos]{todonotes}
\usepackage[version=3]{mhchem}
\usepackage{chemfig}
\usepackage{cancel}
\usepackage{hyperref}
\usepackage{tabularx}
\usepackage{color}
\usepackage{harpoon}
\usepackage{upgreek}
\usepackage{epstopdf}
\usepackage{bm}

\usepackage{changes}

\makeatletter
\newsavebox{\@brx}
\newcommand{\llangle}[1][]{\savebox{\@brx}{\(\m@th{#1\langle}\)}%
	\mathopen{\copy\@brx\kern-0.5\wd\@brx\usebox{\@brx}}}
\newcommand{\rrangle}[1][]{\savebox{\@brx}{\(\m@th{#1\rangle}\)}%
	\mathclose{\copy\@brx\kern-0.5\wd\@brx\usebox{\@brx}}}
\makeatother

\begin{document}
	
	\title{Nonlinear transport fingerprints of tunable Fermi-arc connectivity in magnetic Weyl semimetal Co$_3$Sn$_2$S$_2$
	}
	
	\author{K. X. Jia}
	\affiliation{Interdisciplinary Center for Theoretical Physics and Information Sciences (ICTPIS), Fudan University, Shanghai 200433, China}
	
	\author{H. C. Li}
	\affiliation{
		School of Materials and Physics, China University of Mining and Technology, Xuzhou 221116, China}
	
	\author{M. H. Zou}
	\affiliation{National Laboratory of Solid State Microstructures and Department of
		Physics, Nanjing University, Nanjing 210093, China}
	
	\author{H. Geng}
	\email{genghao@nuaa.edu.cn}
	\affiliation{College of Physics, Nanjing University of Aeronautics and Astronautics, Nanjing 211106, China}
	\affiliation{Key Laboratory of Aerospace Information Materials and Physics (NUAA), MIIT, Nanjing 211106, China}
	
	\author{Hua Jiang}
	\email{jianghuaphy@fudan.edu.cn}
	\affiliation{Interdisciplinary Center for Theoretical Physics and Information Sciences (ICTPIS), Fudan University, Shanghai, 200433, China}

	\date{\today}

	\begin{abstract}
		Fermi arcs in Weyl semimetals provide a unique platform for surface-state engineering, yet directly tracking of their evolution under surface tuning remains experimentally challenging. Here we theoretically propose that nonreciprocal charge transport can serve as a direct probe of Fermi arc Lifshitz transitions (FALT). We show that different surface terminations in Co$_3$Sn$_2$S$_2$ can produce finite and highly tunable second-order nonreciprocal signals, which can be further modulated by adjusting the surface potential. Strikingly, we show that the second-order conductivity exhibits sign changes as the Fermi arc connectivity is tuned across FALT driven by gating or chemical potential variation. This behavior arises from the chiral nature of electron velocities on the Fermi arcs, and is highly sensitive to surface termination and symmetry breaking. Our findings establish nonreciprocal transport as an electrically measurable fingerprint of FALT and propose new strategies that could be directly applied in devices for in situ engineering and detecting transport properties in topological materials.
	\end{abstract}
	
	\maketitle	
	
	\textit{Introduction}.$-$Topological semimetals have reshaped our understanding of gapless quantum matter by hosting exotic surface states that go beyond conventional Fermi liquid theory\cite{Armitage2018}. In Weyl semimetals (WSM), these appear as open Fermi arcs (FAs)—unusual surface bands connecting projections of bulk Weyl points (WPs) with opposite chirality\cite{Wan2011}. While the essential band topology of Weyl systems is well established, recent attention has shifted to the remarkable sensitivity and tunability of their surface states\cite{BedoyaPinto2021,Ekahana2020,Morali2019,Souma2016,Sun2015,Zheng2023,Chaou2024,Lim2024,Kong2024,Chen2024,Ji2023,Mazzola2023,Huang2023,Ovalle2022,Hu2022,Chen2022,Buccheri2022,Zheng2021,Wang2021a,Murthy2020,Shvetsov2020}. Experimental and theoretical advances now make it possible to engineer the connectivity, energy, and symmetry of FAs through surface termination\cite{Huang2023,Morali2019}, gating\cite{Zheng2023}, and surface doping\cite{BedoyaPinto2021}. This surface-specific control enables Fermi arc Lifshitz transitions (FALT), where FAs connectivity can be switched without altering bulk WPs, providing a controllable route for topological reconstruction and device design\cite{Yang2019,Quirk2023,Cheng2024,Wu2023,Hu2022,Wang2022,Zheng2021,Wadge2022}.
	
	The experimental observation of FAs is intrinsically challenging\cite{Xu2015a,Lv2015a,Lv2021,Bernevig2022,SilvaNeto2019}, as their connectivity and detailed structure are highly sensitive to surface quality, termination, and external perturbations. This challenge becomes more pronounced when one aims to track Fermi-arc evolution in situ during surface-state tuning under realistic FAs engineering conditions\cite{BedoyaPinto2021}. As a result, identifying reliable transport signatures that sensitively reflect the presence and reconstruction of FAs is of critical importance for both fundamental studies and future device applications\cite{Armitage2018,Moll2016,Huang2016,Chen2020,Miyazaki2022,Li2020}.
	
	Recent advances in nonreciprocal charge transport (NCT)\cite{Tokura2018,Ideue2017,Itahashi2020, Li2021a, Yasuda2019,Choe2019, Shim2022, Ye2022,Yasuda2020,Zhang2022,Wang2020a,Zeng2021,Das2022} have opened a viable route to address this challenge. NCT refers to the asymmetric charge transport of a system to external fields of opposite directions\cite{Rikken2001, Krstic2002}. 
	Beyond its potential for next‐generation diode‐like functionalities, NCT also offers a highly sensitive probe of symmetry breaking—capable of revealing phase transitions\cite{Wang2022a}, hidden orders\cite{Guo2022}, Berry curvature effects\cite{Ma2018,Wang2023a}, and spin-momentum locking in topological insulators\cite{Yasuda2016}. 
	 
	 In our paper, we present a theoretical investigation of NCT arising from asymmetric FAs. By systematically exploring the dependence of nonreciprocal resistance on temperature, crystal orientation, and chemical potential, we provide a comprehensive characterization of its tunability. We predict a new type of NCT signature directly linked to FALT. We demonstrate that the second-order nonlinear conductivity undergoes sign changes as the FAs connectivity is tuned across critical points, either by gating or by varying the chemical potential. This correspondence between nonlinear transport and Fermi-arc topology provides a practical and experimentally accessible probe of FALT in WSMs and highlights the potential of surface in situ engineering and detecting topological phenomena in quantum materials.
	
	\begin{figure}
		\centering
		\includegraphics[width=1.05\columnwidth]{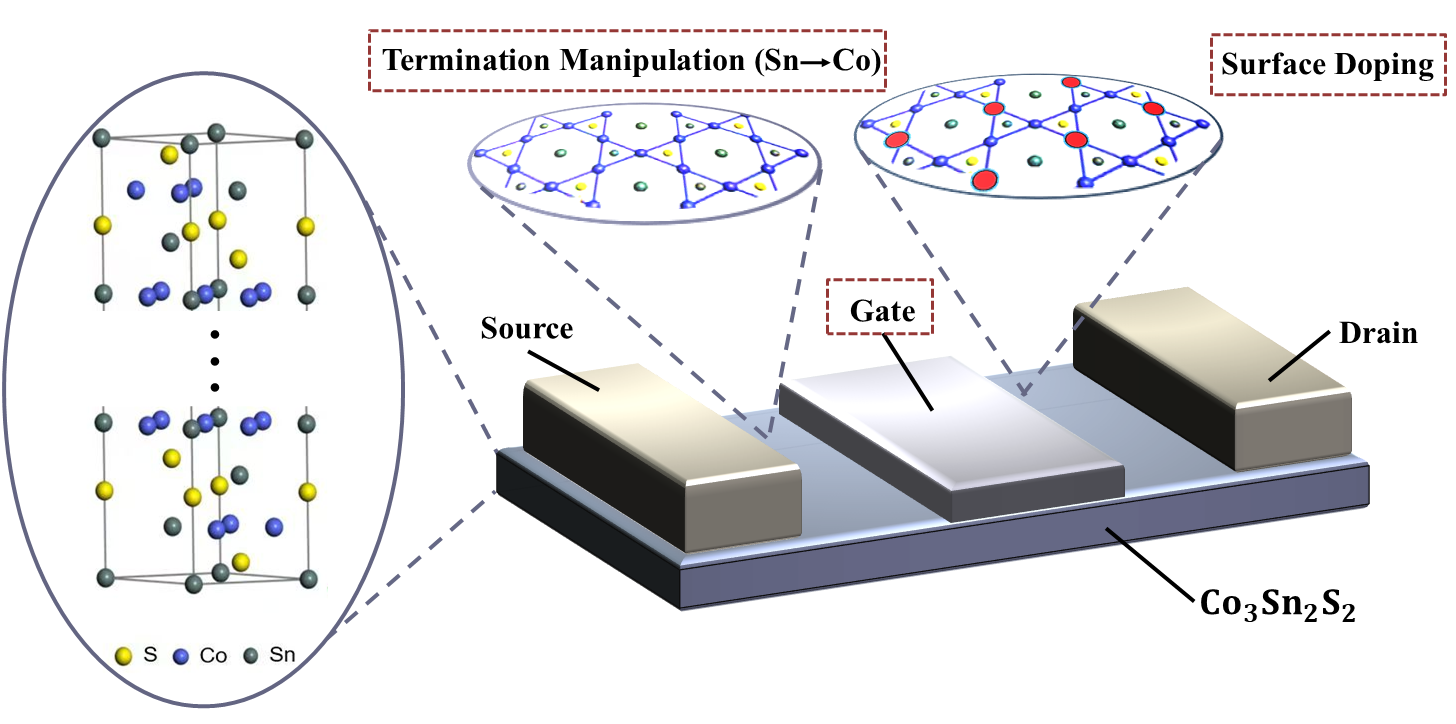}
		\caption{Schematic illustration of surface engineering strategies in Co$_3$Sn$_2$S$_2$ for controlling NCT. The device is based on a Co$_3$Sn$_2$S$_2$ single crystal, where NCT is tuned via three distinct surface engineering approaches: (i) surface termination manipulation (by substituting Sn-terminated surfaces with Co-terminated ones), (ii) surface chemical doping, and (iii) application of an external surface potential (via gating). The atomic structure of Co$_3$Sn$_2$S$_2$ is shown on the left, highlighting the stacking of S, Co, and Sn atoms.
		}\label{fig1}
	\end{figure}
	
	\textit{Model Hamiltonian}.$-$NCT requires the simultaneous breaking of time-reversal and inversion symmetries\cite{Tokura2018}. Here we focus on the magnetic Weyl semimetal Co$_3$Sn$_2$S$_2$, which features a layered kagome lattice, a robust Weyl phase, and readily tunable surface terminations.~\cite{Liu2019a,Huang2023,Howard2021} It has recently emerged as a prominent platform for exploring unconventional transport phenomena\cite{Liu2018,Wang2022b,Yang2025}, and ARPES measurements have revealed that different terminations produce strikingly distinct Fermi-arc connectivities\cite{Morali2019}. The diversity of Fermi arcs together with the ferromagnetic order respectively break inversion and time-reversal symmetries in Co$_3$Sn$_2$S$_2$, making it an ideal platform for investigating nonreciprocal signals.
	
	To investigate the longitudinal transport properties associated with the Fermi-arc surface states of Co$_3$Sn$_2$S$_2$, we construct an effective tight-binding model . Following the minimal model introduced by Ozawa et al.\cite{Ozawa2019}, we consider one \textit{d}-orbital from each Co atom forming the kagome plane and one \textit{p}-orbital from the interlayer Sn atoms, incorporating spin-dependent hopping and Kane-Mele type spin-orbit coupling (SOC). Specifically, the Hamiltonian is given by
	\begin{equation}
		H = H_{\mathrm{hop}} + H_{\mathrm{KM}} + H_{\mathrm{exc}} + H_{\mathrm{surf}},
	\end{equation}
	with
	\begin{equation}
		\begin{split}
		H_{\mathrm{hop}} &= \sum_{ i, j, s} t_{ij}\, d^{\dagger}_{i s} d_{j s} + \
		+\sum_{ i, s} \epsilon_d \, d^{\dagger}_{i s} d_{i s}\\
		&+\sum_{\langle i, j \rangle, s} t'_{ij}\, d^{\dagger}_{i s} p_{j s} + \sum_{ i, s} \epsilon_p \, p^{\dagger}_{i s} p_{i s}+\mathrm{H.c.} \\[0.8em]
		H_{\mathrm{KM}}  &= i \lambda_{\mathrm{KM}} \sum_{\langle\!\langle i, j \rangle\!\rangle, s, s'} \nu_{ij}\, d^{\dagger}_{i s} \sigma^z_{s s'} d_{j s'} \\[0.8em]
		H_{\mathrm{exc}} &= -m \sum_{i, s, s'} d^{\dagger}_{i s} \, \sigma_z \, d_{i s'} \\[0.8em]
		H_{\mathrm{Top}} &= \sum_{i \in \mathrm{Top}, s} (V_{\mathrm{eff}}+V_G) \, d^{\dagger}_{i s} d_{i s},
	    \end{split}
	\end{equation}
	
	where $t_{ij}$ includes the nearest-neighbor hopping $t_1$, the second-nearest-neighbor hopping $t_2$ in the kagome layer and the interlayer kagome hopping $t_z$. $s,s'$ represents the spin degree of freedom. $H_{\mathrm{KM}}$ describes the intralayer Kane-Mele SOC. $H_{\mathrm{exc}}$ represents the exchange interaction where the magnetization is aligned along the $z$ axis. 
	The core principle of FAs engineering is to tune the effective surface potential. In Fig~\ref{fig1}, we illustrated three representative approaches: termination control, surface doping, and gate modulation. For clarity, this study mainly discusses nonreciprocal signals resulting from termination engineering and gating. To capture the variation of the local potential experienced by surface atoms, we further introduce an on-site potential correction $H_{\text{surf}}$ at the surface layer. Specifically, $V_{\text{eff}}$ 
	is the additional uncompensated local field on the top Co atoms when the top and bottom terminations differ due to the absence of S/Sn atoms on the vacuum side. (for identical terminations, $V_{\text{eff}}$ is set to zero). $V_G$ stands for the top gate voltage which is used to drive FALT.
	\begin{figure}
		\centering
		\includegraphics[width=1.05\columnwidth]{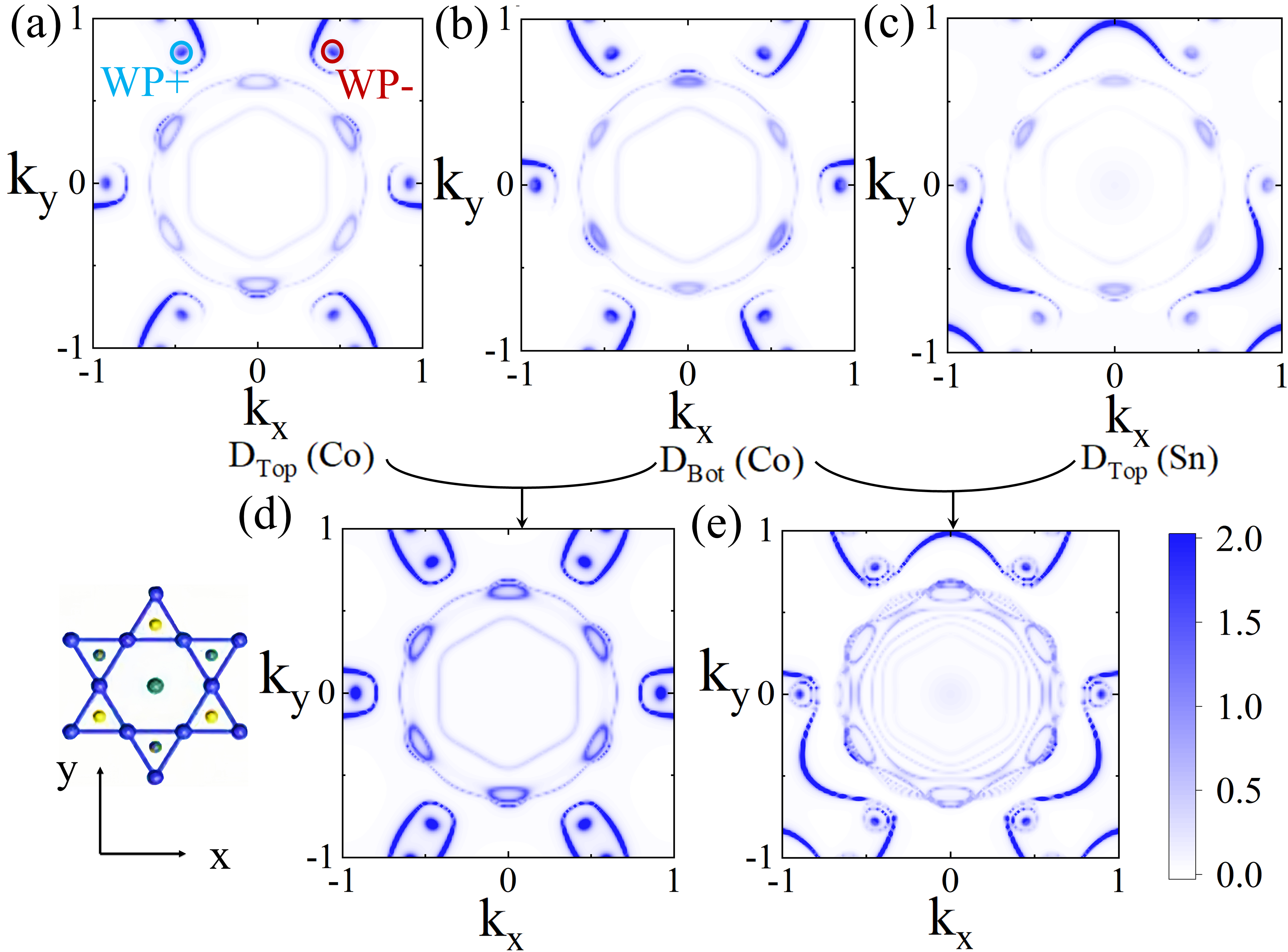}
		\caption{Calculated surface LDOS for various surface terminations of the Co$_3$Sn$_2$S$_2$ slab at $\mu=0.18eV$. Panels~(a) and (b) correspond to Co-terminated top and bottom surfaces, respectively, while panel~(c) shows the result for the Sn-terminated top surface. (d, e) FAs connectivity for slabs with symmetric (Co–Co) and asymmetric (Co–Sn) surface terminations. 
		}\label{fig2}
	\end{figure}

	To study the spectral features of the arc states, we impose open boundary conditions along the $z$ direction and perform Fourier transformation in the remaining two in-plane directions. In this mixed representation, the retarded Green's function is given by
	\begin{equation}
		G^{s,s'}(z,z',\mathbf{k},\omega)=\langle z,\mathbf{k},s|\frac{1}{\omega-H+i0^{+}}|z',\mathbf{k},s'\rangle.
	\end{equation}
	The local density of states (LDOS) for the surface states is then obtained from the spectral function
	\begin{equation}
		A(\mathbf{k},\omega)=-\mathrm{Im}\sum_{s,z\in z_{S}} G^{s,s}(z,z,\mathbf{k},\omega)/\pi,
	\end{equation}
	where for each surface termination considered, we select the outermost Co and Sn atomic layers $(z\in z_{S})$ as the representative surface sites.
	
	When both surfaces are terminated by Co atoms, the FAs on the two opposite surfaces are related by spatial inversion symmetry and collectively form a set of closed, symmetric Weyl orbits connecting projections of bulk WPs as shown in Fig.~\ref{fig2} (d). In contrast, when the slab has different terminations (Co-Sn), the connectivity of the surface FAs is altered:  asymmetric terminations reduce the $C_6$ symmetry to $C_3$ symmetry, and the FAs on both surfaces are no longer related by inversion. Specifically, the Fermi surface is symmetric about $k_x = 0$, but asymmetric about $k_y = 0$, as shown in Fig.~\ref{fig2} (e). 
	Note that this asymmetry in surface-state connectivity becomes particularly significant when the Fermi energy is tuned close to the WPs, where the bulk density of states is intrinsically suppressed due to the linear band dispersion. In this regime, the noncentro-symmetric Fermi-arc connectivity between different terminations gives rise to robust NCT phenomena. 
	
	\begin{figure*}[!t]
		\centering
		\includegraphics[width=2.0\columnwidth]{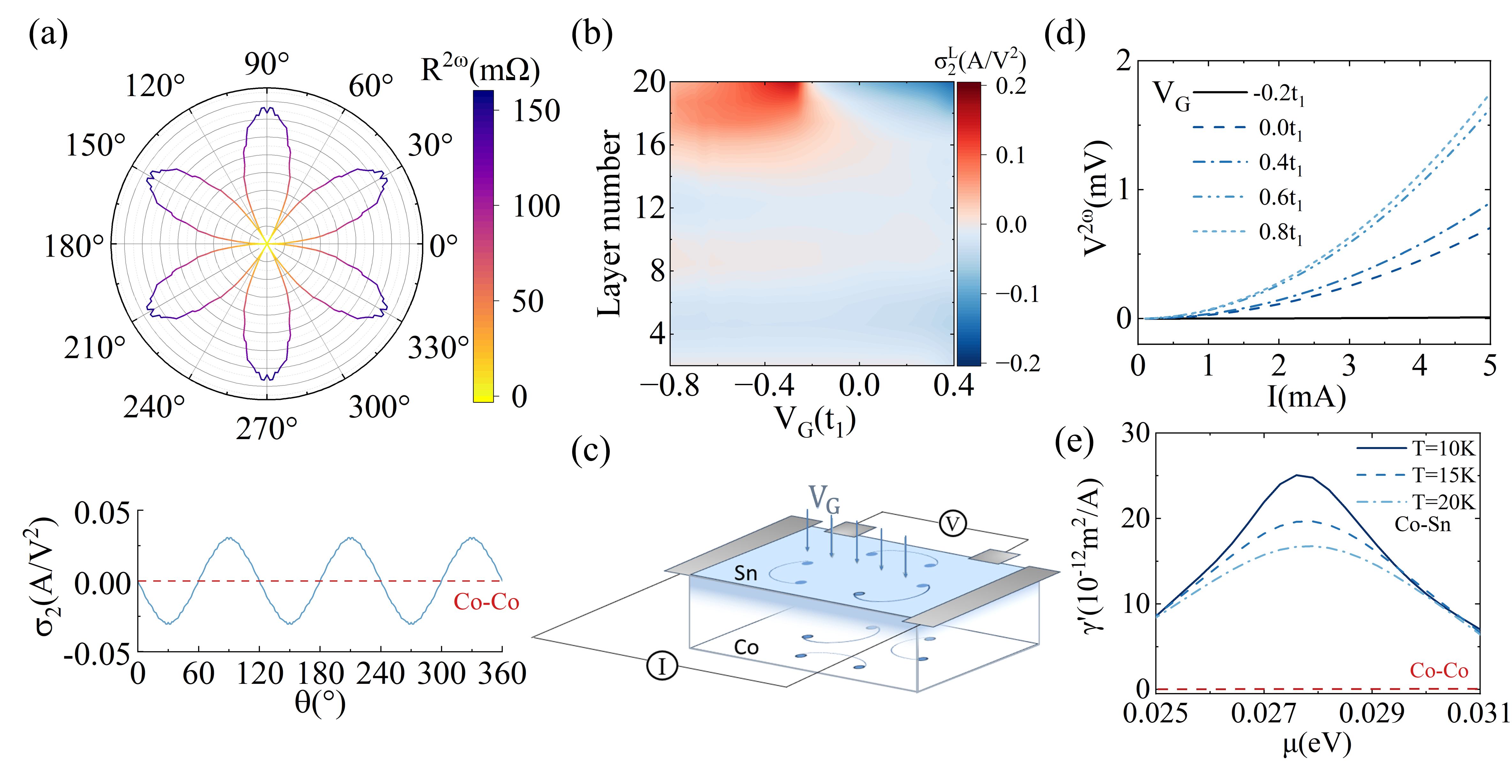}
		\caption{Anisotropic and tunable second-order nonreciprocal response in Co$_3$Sn$_2$S$_2$.
			(a) Angular dependence of the second-order nonlinear conductivity $\sigma_2$, calculated by numerical integration of Eq.~(8) after coordinate rotation.
			(b) Layer-resolved contribution to the second-order conductivity at $\mu=0.18eV$, which can be tuned by applying a gate voltage to the Sn-terminated surface in (c).
			(d) Gate-controlled enhancement of the second-harmonic voltage $V^{2\omega}$ with increasing surface potential $V_{\mathrm{G}}$.
			(e) Nonreciprocal coefficient $\gamma’$ versus chemical potential $\mu$ for different temperatures and surface terminations. The signal peaks near the WPs for Co–Sn surfaces, but vanishes for symmetric Co–Co cases.
		}\label{fig3}
	\end{figure*}

	\textit{Semiclassical Boltzmann approach to nonlinear conductivity.}$-$To elucidate the microscopic origin of the second-order nonlinear conductivity, we adopt a semiclassical Boltzmann formalism under an external electric field. The Boltzmann equation for the distribution function $f$ reads
	\begin{equation}
		\partial_t f+\frac{d\mathbf{k}}{dt} \cdot \nabla_{\mathbf{k}} f = -\frac{f - f_0}{\tau},
	\end{equation}
	where $f_0$ is the equilibrium Fermi-Dirac distribution, $\tau$ is the relaxation time. The nonequilibrium distribution can be expanded as $f = f_0 + f_1 + f_2 + \cdots$, with $f_n$ describing the $n$th-order correction of 
	$\mathbf{E}$. Here we focus on the longitudinal current along the y-direction. Explicitly, 
	\begin{align}
		f_1=\frac{eE\tau}{\hbar}\frac{\partial f_0}{\partial k_y},\quad
		f_2=\tau^2 e^2 E^2\left(\frac{\partial v_y}{\hbar\partial k_y}\frac{\partial f_0}{\partial \varepsilon}+v_y^2\frac{\partial^2 f_0}{\partial^2 \varepsilon}\right),
	\end{align}
	where $\varepsilon(n,\mathbf{k})$ is the energy and $v_y=\frac{\partial \varepsilon}{\partial k_y}$ is the velocity. The resulting current is then given by
	\begin{equation}
		j_y = e \sum_n\int v_{y} f_2 \, d{\mathbf{k}} + e^2 E^2  \sum_n\int \frac{\partial G^{yy}_n}{\hbar\partial \mathbf{k}} f_0 \, d\mathbf{k},
	\end{equation}
	where the first term corresponds to the Drude contribution, while the second term arises from the quantum metric correction to the group velocity\cite{Kaplan2024a}. Here, $G^{yy}_n = \sum_{m \neq n} \mathrm{Re}\left(\frac{A^y_{nm} A^y_{mn}}{\varepsilon_m - \varepsilon_n}\right)$ encodes interband coherence, with $A^y_{nm} = \langle n\mathbf{k} | \frac{\partial}{\partial k_y} | m\mathbf{k} \rangle$ the Berry
	connection term. The sum over $n$ denotes all layers along the 
	$z$-direction. The second-order nonlinear conductivity tensor $\sigma_2$ is then given by 
	\begin{align}
		\sigma_2 &= -\frac{e^3 \tau^2}{\hbar^3}\sum_n \int \frac{\partial^3 \varepsilon_n}{\partial k_y^3} f_0 d\mathbf{k} 
		- \frac{e^3}{\hbar} \sum_n\int \frac{\partial G^{yy}_n}{\partial k_y}  f_0 d\mathbf{k}.
	\end{align}
	Note that the contribution of the quantum metric dipole term to $\sigma_2$ is more than one order of magnitude weaker than Drude term (see Appendix for details). As a result, in the following analysis we focus exclusively on the Drude contribution to $\sigma_2$. To further disentangle the contributions of different surfaces and the bulk to the nonreciprocal signal, we define a layer-resolved nonlinear conductivity
	\begin{align}
		\sigma^L_2(z)= &= -\frac{e^3 \tau^2}{\hbar^3}\sum_n \int \langle\psi_{\mathbf{k},n}|\frac{\partial^3 \varepsilon_n}{\partial k_y^3} f_0| \psi_{z}\rangle\langle \psi_{z} | \psi_{\mathbf{k},n}\rangle d\mathbf{k} 
		,
	\end{align}
	where $\psi_{\mathbf{k},n}$ are the energy eigenstates and $| \psi_{z}\rangle\langle \psi_{z} |$ is the projecting operator.
	 We use the following parameters unless otherwise specified\cite{Ozawa2019}: $t_1=0.15eV,~t_2=0.6t_1,~t'=t_1,~t_3=-t_1,~\lambda_{KM}=0.3t_1,~\epsilon_p=-1.5t_1,~\epsilon_d=2.0t_1,~m=2.0t_1,~V_{\text{eff}}=1.2t_1$.
	
	The second-order nonlinear conductivity in Co$_3$Sn$_2$S$_2$ is governed by the third-rank conductivity tensor $\sigma^{(2)}_{ijk}$. In our system, $\sigma^{(2)}_{ijk}$ possesses only a single independent component, $\sigma \equiv \sigma^{(2)}_{yyy}$, with all other nonzero components related by symmetry as $\sigma^{(2)}_{yyy} = -\sigma^{(2)}_{yxx} = -\sigma^{(2)}_{xxy} = -\sigma^{(2)}_{xyx}$ (see Appendix for details). When an in-plane electric field $\mathbf{E} = E(\cos\theta, \sin\theta)$ is applied, the second-order nonlinear current along the direction $\mathbf{n} = (\cos\theta, \sin\theta)$ is given by
	\begin{equation}
		j_n^{(2)} = \sigma E^2 \sin(3\theta)
		\label{eq:C3_second_order}
	\end{equation}
	where $\theta$ is the angle relative to the crystal $x$ axis. This $\sin(3\theta)$ dependence\cite{Ma2018,Kaplan2024} gives rise to an angular dependence in the nonlinear response as the field orientation $\theta$ is rotated in the $xy$ plane, which agrees well with the numerical calculations presented in Fig.~\ref{fig3} (a).
	
	To illustrate a feasible detection scheme for the nonreciprocal surface signal, we model a Co$_3$Sn$_2$S$_2$ thin film consisting of $L_z=20$ Co layers—sufficient to retain the Weyl phase\cite{Ikeda2021a} while keeping the computational cost manageable. The lateral dimensions are set to $L_x=10\mu m,~L_y=10\mu m$, and a gate voltage $V_G$ is applied to the top surface to modulate the surface potential\cite{Zheng2021}, as shown in Fig.~\ref{fig3} (c).
	
	To quantitatively characterize the nonlinear transport, we define the nonreciprocal coefficient $\gamma$ as $R =R_0(1+\gamma I)$ (see Appendix for details) with $\gamma=-\frac{1}{L_xL_z}\frac{\sigma_2}{\sigma_1^2}$ and $R_0$ the resistance along the x direction. We further define a dimension-independent quantity as $\gamma’=L_xL_z\gamma$ to provides a direct measure of the strength of the nonreciprocal effect. As illustrated in Fig.~\ref{fig3}(e), $\gamma'$ shows a peak when the chemical potential $\mu$ approaches the WPs. This enhancement occurs because the bulk density of states is suppressed near the WPs, allowing FAs to dominate transport. As  $T$ increases, thermal broadening reduces the sharpness and magnitude of the peak. Nevertheless, owing to the high Curie temperature of Co$_3$Sn$_2$S$_2$ ($T_{\mathrm{C}} \sim 175~\mathrm{K}$), the nonreciprocal signal remains observable over a broad $T$ range. Notably, when both surfaces of the slab are terminated identically (Co–Co configuration), the system preserves inversion symmetry between top and bottom surfaces, resulting in a vanishing nonreciprocal response, as shown in Fig.~\ref{fig3} (e).
	
	The nonreciprocal response can also be tuned by engineering the surface potential through gating. Adjusting the effective surface potential $V_{G}$ effectively shifts the Fermi energy $\mu$ of the FAs, thereby enabling control over the amplitude of the nonreciprocal signal. This mechanism is demonstrated in Fig.~\ref{fig3}(d), where the second-order voltage $V^{2\omega}$ exhibits clear quadratic scaling with the fundamental current $I$, and its magnitude increases as $V_{G}$ is raised. Collectively, these results highlight the tunability of the nonreciprocal transport.  Note that the relative contributions from different surfaces are highly sensitive to the effective surface potential. As shown in Fig.~\ref{fig3}(b), under Sn–Co termination, the top surface dominates the response, which underpins our NCT as a probe for FALT discussed below. 

\begin{figure*}
	\centering
	\includegraphics[width=2.0\columnwidth]{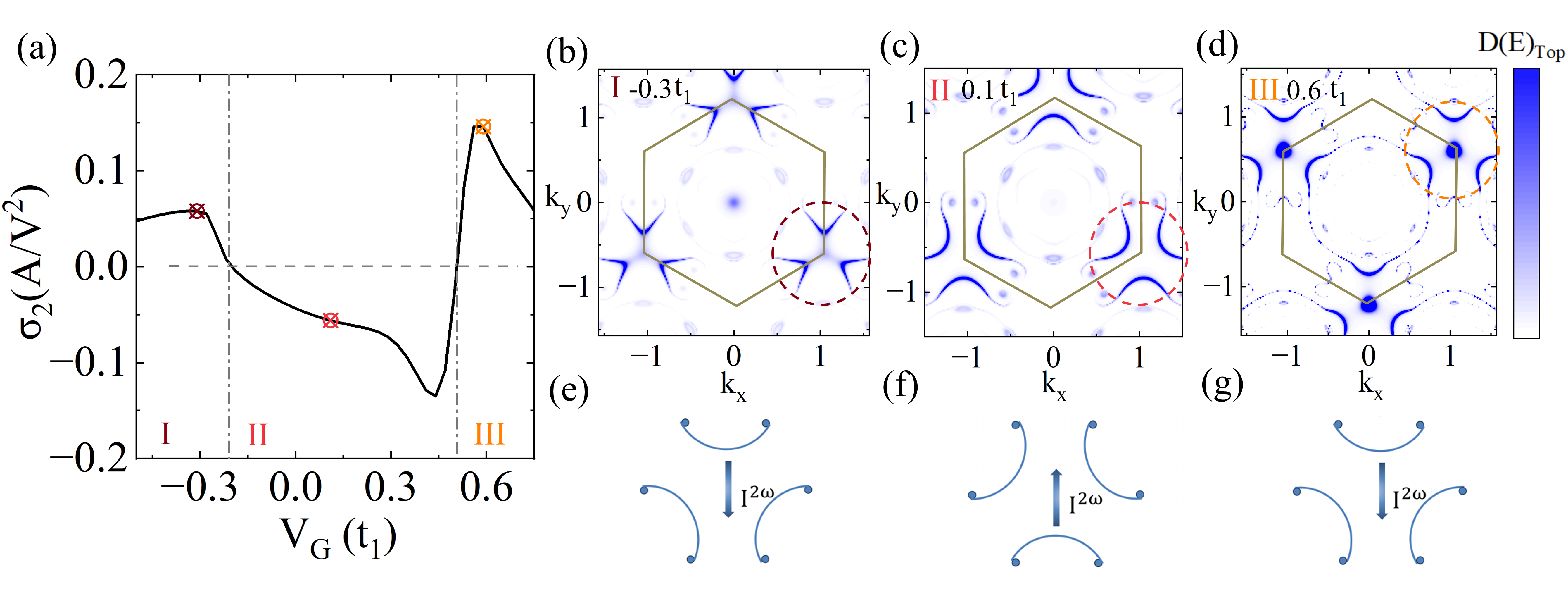}
	\caption{Gate-tunable FALT and nonreciprocal nonlinear transport in Co$_3$Sn$_2$S$_2$.
		(a) Calculated second-order conductivity $\sigma_2$ as a function of the gate voltage $V_\mathrm{G}$. Distinct regimes (I, II, III) are separated by critical values of $V_\mathrm{G}$, corresponding to topological FALT.
		(b-d) Momentum-resolved surface LDOS on the Sn-terminated surface for representative values of $V_\mathrm{G}$. The FAs connectivity undergoes characteristic reconstructions across the three regimes, with sign changes at the transition points.
	}\label{fig4}
\end{figure*}
	
	\textit{NCT as a probe of FALT}.$-$As a manifestation of the nonreciprocal transport, a central result of this work is the identification of a new class of transport signatures for FALT. As plotted in Fig.~\ref{fig4}(b), as $V_\mathrm{G}$ tuned, the connectivity of the FAs undergoes a sequence of Lifshitz transitions, with characteristic changes in their shapes and connectivity patterns. Remarkably, each transition is accompanied by a sign reversal of $\sigma_2$ as shown in Fig.~\ref{fig4}(a), indicating a strong sensitivity of the NCT to the underlying FAs topology.  
	
	The physical mechanism underlying the abrupt switching of the second-order nonreciprocal current $I^{2\omega}$ is rooted in the chiral nature of electron velocities on the FAs. 
	The three modes in Fig. ~\ref{fig4} can be understood through Fig.~\ref{fig4} (e-g). Among them, modes I and III have opposite chirality to mode II, which, according to Eq. (12), is equivalent to rotating the measurement direction by $180°$, and thus produces an opposite nonreciprocal signal. Furthermore, near the Lifshitz transition point the second-order response vanishes, since the contributions from two opposite FAs structures mutually cancel(see Fig.~\ref{fig6} in Appendix for details).
	
	It should be noted that our theoretical analysis has focused on the FAs structure of the top surface. In practical measurements, additional contributions from the bottom surface may also appear. Nevertheless, as discussed earlier, for the Sn–Co termination the dominant response originates from the Sn-terminated surface. In addition, the top surface is naturally more accessible for electrical contacts in real devices, offering practical advantages for detecting the predicted second-order conductivity and verifying its surface origin.
	
	\textit{Conclusion}.$-$In conclusion, we have theoretically uncovered that different surface terminations in WSMs can induce a finite and tunable second-order nonreciprocal conductivity, which can be efficiently controlled via surface potential engineering. By systematically analyzing Fermi arc–driven surface Lifshitz transition in Co$_3$Sn$_2$S$_2$, we find that abrupt changes in the topology of Fermi arcs lead to sign reversals and strong modulation of the nonlinear signal. These findings bridge the gap between topological surface-state manipulation and measurable electrical responses, opening new avenues for the in situ control and measurement of topological states and related exotic transport phenomena under realistic working conditions. We anticipate that our predictions can be tested in Co$_3$Sn$_2$S$_2$ devices, and that may be extended to a wide class of topological materials such as WSM NbAs\cite{Yang2019}, Co$_2$MnGa\cite{Belopolski2019} and magnetic topological insulator MnBi$_2$Te$_4$.
	
	\begin{acknowledgments}
		This work was supported by  the National Natural Science Foundation of China under Grant
		No. 12304068 (H. G.),
		No. 12274235 (R.M.),
		No. 12174182 (D.Y.X.),
		the startup Fund of Nanjing University of Aeronautics and Astronautics Grant No. YAH24076 (H. G.),
		and the State Key Program for Basic Researches of
		China under Grants No. 2021YFA1400403 (D.Y.X.). 
		The computations are partially supported by High Performance Computing Platform of Nanjing University of Aeronautics and Astronautics.
	\end{acknowledgments}
	

\begin{thebibliography}{69}%
\makeatletter
\providecommand \@ifxundefined [1]{%
 \@ifx{#1\undefined}
}%
\providecommand \@ifnum [1]{%
 \ifnum #1\expandafter \@firstoftwo
 \else \expandafter \@secondoftwo
 \fi
}%
\providecommand \@ifx [1]{%
 \ifx #1\expandafter \@firstoftwo
 \else \expandafter \@secondoftwo
 \fi
}%
\providecommand \natexlab [1]{#1}%
\providecommand \enquote  [1]{``#1''}%
\providecommand \bibnamefont  [1]{#1}%
\providecommand \bibfnamefont [1]{#1}%
\providecommand \citenamefont [1]{#1}%
\providecommand \href@noop [0]{\@secondoftwo}%
\providecommand \href [0]{\begingroup \@sanitize@url \@href}%
\providecommand \@href[1]{\@@startlink{#1}\@@href}%
\providecommand \@@href[1]{\endgroup#1\@@endlink}%
\providecommand \@sanitize@url [0]{\catcode `\\12\catcode `\$12\catcode
  `\&12\catcode `\#12\catcode `\^12\catcode `\_12\catcode `\%12\relax}%
\providecommand \@@startlink[1]{}%
\providecommand \@@endlink[0]{}%
\providecommand \url  [0]{\begingroup\@sanitize@url \@url }%
\providecommand \@url [1]{\endgroup\@href {#1}{\urlprefix }}%
\providecommand \urlprefix  [0]{URL }%
\providecommand \Eprint [0]{\href }%
\providecommand \doibase [0]{https://doi.org/}%
\providecommand \selectlanguage [0]{\@gobble}%
\providecommand \bibinfo  [0]{\@secondoftwo}%
\providecommand \bibfield  [0]{\@secondoftwo}%
\providecommand \translation [1]{[#1]}%
\providecommand \BibitemOpen [0]{}%
\providecommand \bibitemStop [0]{}%
\providecommand \bibitemNoStop [0]{.\EOS\space}%
\providecommand \EOS [0]{\spacefactor3000\relax}%
\providecommand \BibitemShut  [1]{\csname bibitem#1\endcsname}%
\let\auto@bib@innerbib\@empty
\bibitem [{\citenamefont {Armitage}\ \emph {et~al.}(2018)\citenamefont
  {Armitage}, \citenamefont {Mele},\ and\ \citenamefont
  {Vishwanath}}]{Armitage2018}%
  \BibitemOpen
  \bibfield  {author} {\bibinfo {author} {\bibfnamefont {N.~P.}\ \bibnamefont
  {Armitage}}, \bibinfo {author} {\bibfnamefont {E.~J.}\ \bibnamefont {Mele}},\
  and\ \bibinfo {author} {\bibfnamefont {A.}~\bibnamefont {Vishwanath}},\
  }\href {https://doi.org/10.1103/RevModPhys.90.015001} {\bibfield  {journal}
  {\bibinfo  {journal} {Rev. Mod. Phys.}\ }\textbf {\bibinfo {volume} {90}},\
  \bibinfo {pages} {015001} (\bibinfo {year} {2018})}\BibitemShut {NoStop}%
\bibitem [{\citenamefont {Wan}\ \emph {et~al.}(2011)\citenamefont {Wan},
  \citenamefont {Turner}, \citenamefont {Vishwanath},\ and\ \citenamefont
  {Savrasov}}]{Wan2011}%
  \BibitemOpen
  \bibfield  {author} {\bibinfo {author} {\bibfnamefont {X.}~\bibnamefont
  {Wan}}, \bibinfo {author} {\bibfnamefont {A.~M.}\ \bibnamefont {Turner}},
  \bibinfo {author} {\bibfnamefont {A.}~\bibnamefont {Vishwanath}},\ and\
  \bibinfo {author} {\bibfnamefont {S.~Y.}\ \bibnamefont {Savrasov}},\ }\href
  {https://doi.org/10.1103/physrevb.83.205101} {\bibfield  {journal} {\bibinfo
  {journal} {Physical Review B}\ }\textbf {\bibinfo {volume} {83}},\ \bibinfo
  {pages} {205101} (\bibinfo {year} {2011})}\BibitemShut {NoStop}%
\bibitem [{\citenamefont {Bedoya-Pinto}\ \emph {et~al.}(2021)\citenamefont
  {Bedoya-Pinto}, \citenamefont {Liu}, \citenamefont {Tan}, \citenamefont
  {Pandeya}, \citenamefont {Chang}, \citenamefont {Zhang},\ and\ \citenamefont
  {Parkin}}]{BedoyaPinto2021}%
  \BibitemOpen
  \bibfield  {author} {\bibinfo {author} {\bibfnamefont {A.}~\bibnamefont
  {Bedoya-Pinto}}, \bibinfo {author} {\bibfnamefont {D.}~\bibnamefont {Liu}},
  \bibinfo {author} {\bibfnamefont {H.}~\bibnamefont {Tan}}, \bibinfo {author}
  {\bibfnamefont {A.~K.}\ \bibnamefont {Pandeya}}, \bibinfo {author}
  {\bibfnamefont {K.}~\bibnamefont {Chang}}, \bibinfo {author} {\bibfnamefont
  {J.}~\bibnamefont {Zhang}},\ and\ \bibinfo {author} {\bibfnamefont
  {S.~S.~P.}\ \bibnamefont {Parkin}},\ }\href
  {https://doi.org/10.1002/adma.202008634} {\bibfield  {journal} {\bibinfo
  {journal} {Advanced Materials}\ }\textbf {\bibinfo {volume} {33}},\ \bibinfo
  {pages} {2008634} (\bibinfo {year} {2021})}\BibitemShut {NoStop}%
\bibitem [{\citenamefont {Ekahana}\ \emph {et~al.}(2020)\citenamefont
  {Ekahana}, \citenamefont {Li}, \citenamefont {Sun}, \citenamefont {Namiki},
  \citenamefont {Yang}, \citenamefont {Jiang}, \citenamefont {Yang},
  \citenamefont {Shi}, \citenamefont {Zhang}, \citenamefont {Pei},
  \citenamefont {Chen}, \citenamefont {Sasagawa}, \citenamefont {Felser},
  \citenamefont {Yan}, \citenamefont {Liu},\ and\ \citenamefont
  {Chen}}]{Ekahana2020}%
  \BibitemOpen
  \bibfield  {author} {\bibinfo {author} {\bibfnamefont {S.~A.}\ \bibnamefont
  {Ekahana}}, \bibinfo {author} {\bibfnamefont {Y.~W.}\ \bibnamefont {Li}},
  \bibinfo {author} {\bibfnamefont {Y.}~\bibnamefont {Sun}}, \bibinfo {author}
  {\bibfnamefont {H.}~\bibnamefont {Namiki}}, \bibinfo {author} {\bibfnamefont
  {H.~F.}\ \bibnamefont {Yang}}, \bibinfo {author} {\bibfnamefont
  {J.}~\bibnamefont {Jiang}}, \bibinfo {author} {\bibfnamefont {L.~X.}\
  \bibnamefont {Yang}}, \bibinfo {author} {\bibfnamefont {W.~J.}\ \bibnamefont
  {Shi}}, \bibinfo {author} {\bibfnamefont {C.~F.}\ \bibnamefont {Zhang}},
  \bibinfo {author} {\bibfnamefont {D.}~\bibnamefont {Pei}}, \bibinfo {author}
  {\bibfnamefont {C.}~\bibnamefont {Chen}}, \bibinfo {author} {\bibfnamefont
  {T.}~\bibnamefont {Sasagawa}}, \bibinfo {author} {\bibfnamefont
  {C.}~\bibnamefont {Felser}}, \bibinfo {author} {\bibfnamefont {B.~H.}\
  \bibnamefont {Yan}}, \bibinfo {author} {\bibfnamefont {Z.~K.}\ \bibnamefont
  {Liu}},\ and\ \bibinfo {author} {\bibfnamefont {Y.~L.}\ \bibnamefont
  {Chen}},\ }\href {https://doi.org/10.1103/PhysRevB.102.085126} {\bibfield
  {journal} {\bibinfo  {journal} {Phys. Rev. B}\ }\textbf {\bibinfo {volume}
  {102}},\ \bibinfo {pages} {085126} (\bibinfo {year} {2020})}\BibitemShut
  {NoStop}%
\bibitem [{\citenamefont {Morali}\ \emph {et~al.}(2019)\citenamefont {Morali},
  \citenamefont {Batabyal}, \citenamefont {Nag}, \citenamefont {Liu},
  \citenamefont {Xu}, \citenamefont {Sun}, \citenamefont {Yan}, \citenamefont
  {Felser}, \citenamefont {Avraham},\ and\ \citenamefont
  {Beidenkopf}}]{Morali2019}%
  \BibitemOpen
  \bibfield  {author} {\bibinfo {author} {\bibfnamefont {N.}~\bibnamefont
  {Morali}}, \bibinfo {author} {\bibfnamefont {R.}~\bibnamefont {Batabyal}},
  \bibinfo {author} {\bibfnamefont {P.~K.}\ \bibnamefont {Nag}}, \bibinfo
  {author} {\bibfnamefont {E.}~\bibnamefont {Liu}}, \bibinfo {author}
  {\bibfnamefont {Q.}~\bibnamefont {Xu}}, \bibinfo {author} {\bibfnamefont
  {Y.}~\bibnamefont {Sun}}, \bibinfo {author} {\bibfnamefont {B.}~\bibnamefont
  {Yan}}, \bibinfo {author} {\bibfnamefont {C.}~\bibnamefont {Felser}},
  \bibinfo {author} {\bibfnamefont {N.}~\bibnamefont {Avraham}},\ and\ \bibinfo
  {author} {\bibfnamefont {H.}~\bibnamefont {Beidenkopf}},\ }\href
  {https://doi.org/10.1126/science.aav2334} {\bibfield  {journal} {\bibinfo
  {journal} {Science}\ }\textbf {\bibinfo {volume} {365}},\ \bibinfo {pages}
  {1286} (\bibinfo {year} {2019})}\BibitemShut {NoStop}%
\bibitem [{\citenamefont {Souma}\ \emph {et~al.}(2016)\citenamefont {Souma},
  \citenamefont {Wang}, \citenamefont {Kotaka}, \citenamefont {Sato},
  \citenamefont {Nakayama}, \citenamefont {Tanaka}, \citenamefont {Kimizuka},
  \citenamefont {Takahashi}, \citenamefont {Yamauchi}, \citenamefont {Oguchi},
  \citenamefont {Segawa},\ and\ \citenamefont {Ando}}]{Souma2016}%
  \BibitemOpen
  \bibfield  {author} {\bibinfo {author} {\bibfnamefont {S.}~\bibnamefont
  {Souma}}, \bibinfo {author} {\bibfnamefont {Z.}~\bibnamefont {Wang}},
  \bibinfo {author} {\bibfnamefont {H.}~\bibnamefont {Kotaka}}, \bibinfo
  {author} {\bibfnamefont {T.}~\bibnamefont {Sato}}, \bibinfo {author}
  {\bibfnamefont {K.}~\bibnamefont {Nakayama}}, \bibinfo {author}
  {\bibfnamefont {Y.}~\bibnamefont {Tanaka}}, \bibinfo {author} {\bibfnamefont
  {H.}~\bibnamefont {Kimizuka}}, \bibinfo {author} {\bibfnamefont
  {T.}~\bibnamefont {Takahashi}}, \bibinfo {author} {\bibfnamefont
  {K.}~\bibnamefont {Yamauchi}}, \bibinfo {author} {\bibfnamefont
  {T.}~\bibnamefont {Oguchi}}, \bibinfo {author} {\bibfnamefont
  {K.}~\bibnamefont {Segawa}},\ and\ \bibinfo {author} {\bibfnamefont
  {Y.}~\bibnamefont {Ando}},\ }\href
  {https://doi.org/10.1103/PhysRevB.93.161112} {\bibfield  {journal} {\bibinfo
  {journal} {Phys. Rev. B}\ }\textbf {\bibinfo {volume} {93}},\ \bibinfo
  {pages} {161112} (\bibinfo {year} {2016})}\BibitemShut {NoStop}%
\bibitem [{\citenamefont {Sun}\ \emph {et~al.}(2015)\citenamefont {Sun},
  \citenamefont {Wu},\ and\ \citenamefont {Yan}}]{Sun2015}%
  \BibitemOpen
  \bibfield  {author} {\bibinfo {author} {\bibfnamefont {Y.}~\bibnamefont
  {Sun}}, \bibinfo {author} {\bibfnamefont {S.-C.}\ \bibnamefont {Wu}},\ and\
  \bibinfo {author} {\bibfnamefont {B.}~\bibnamefont {Yan}},\ }\href
  {https://doi.org/10.1103/PhysRevB.92.115428} {\bibfield  {journal} {\bibinfo
  {journal} {Phys. Rev. B}\ }\textbf {\bibinfo {volume} {92}},\ \bibinfo
  {pages} {115428} (\bibinfo {year} {2015})}\BibitemShut {NoStop}%
\bibitem [{\citenamefont {Zheng}\ \emph {et~al.}(2023)\citenamefont {Zheng},
  \citenamefont {Chen}, \citenamefont {Wan},\ and\ \citenamefont
  {Xing}}]{Zheng2023}%
  \BibitemOpen
  \bibfield  {author} {\bibinfo {author} {\bibfnamefont {Y.}~\bibnamefont
  {Zheng}}, \bibinfo {author} {\bibfnamefont {W.}~\bibnamefont {Chen}},
  \bibinfo {author} {\bibfnamefont {X.}~\bibnamefont {Wan}},\ and\ \bibinfo
  {author} {\bibfnamefont {D.~Y.}\ \bibnamefont {Xing}},\ }\href
  {https://doi.org/10.1088/0256-307X/40/9/097301} {\bibfield  {journal}
  {\bibinfo  {journal} {Chinese Physics Letters}\ }\textbf {\bibinfo {volume}
  {40}},\ \bibinfo {pages} {097301} (\bibinfo {year} {2023})}\BibitemShut
  {NoStop}%
\bibitem [{\citenamefont {Chaou}\ \emph {et~al.}(2024)\citenamefont {Chaou},
  \citenamefont {Dwivedi},\ and\ \citenamefont {Breitkreiz}}]{Chaou2024}%
  \BibitemOpen
  \bibfield  {author} {\bibinfo {author} {\bibfnamefont {A.~Y.}\ \bibnamefont
  {Chaou}}, \bibinfo {author} {\bibfnamefont {V.}~\bibnamefont {Dwivedi}},\
  and\ \bibinfo {author} {\bibfnamefont {M.}~\bibnamefont {Breitkreiz}},\
  }\href {https://doi.org/10.1103/PhysRevB.110.035116} {\bibfield  {journal}
  {\bibinfo  {journal} {Phys. Rev. B}\ }\textbf {\bibinfo {volume} {110}},\
  \bibinfo {pages} {035116} (\bibinfo {year} {2024})}\BibitemShut {NoStop}%
\bibitem [{\citenamefont {Lim}\ \emph {et~al.}(2024)\citenamefont {Lim},
  \citenamefont {Kim}, \citenamefont {Lim}, \citenamefont {Kim}, \citenamefont
  {Lee}, \citenamefont {Cha}, \citenamefont {Lee}, \citenamefont {Song},
  \citenamefont {Thapa}, \citenamefont {Denlinger}, \citenamefont {Kim},
  \citenamefont {Kim}, \citenamefont {Seo},\ and\ \citenamefont
  {Kim}}]{Lim2024}%
  \BibitemOpen
  \bibfield  {author} {\bibinfo {author} {\bibfnamefont {C.-y.}\ \bibnamefont
  {Lim}}, \bibinfo {author} {\bibfnamefont {M.-S.}\ \bibnamefont {Kim}},
  \bibinfo {author} {\bibfnamefont {D.~C.}\ \bibnamefont {Lim}}, \bibinfo
  {author} {\bibfnamefont {S.}~\bibnamefont {Kim}}, \bibinfo {author}
  {\bibfnamefont {Y.}~\bibnamefont {Lee}}, \bibinfo {author} {\bibfnamefont
  {J.}~\bibnamefont {Cha}}, \bibinfo {author} {\bibfnamefont {G.}~\bibnamefont
  {Lee}}, \bibinfo {author} {\bibfnamefont {S.~Y.}\ \bibnamefont {Song}},
  \bibinfo {author} {\bibfnamefont {D.}~\bibnamefont {Thapa}}, \bibinfo
  {author} {\bibfnamefont {J.~D.}\ \bibnamefont {Denlinger}}, \bibinfo {author}
  {\bibfnamefont {S.-G.}\ \bibnamefont {Kim}}, \bibinfo {author} {\bibfnamefont
  {S.~W.}\ \bibnamefont {Kim}}, \bibinfo {author} {\bibfnamefont
  {J.}~\bibnamefont {Seo}},\ and\ \bibinfo {author} {\bibfnamefont
  {Y.}~\bibnamefont {Kim}},\ }\bibfield  {journal} {\bibinfo  {journal} {NATURE
  COMMUNICATIONS}\ }\textbf {\bibinfo {volume} {15}},\ \href
  {https://doi.org/10.1038/s41467-024-49841-6} {10.1038/s41467-024-49841-6}
  (\bibinfo {year} {2024})\BibitemShut {NoStop}%
\bibitem [{\citenamefont {Kong}\ \emph {et~al.}(2024)\citenamefont {Kong},
  \citenamefont {Tao}, \citenamefont {Zhang}, \citenamefont {Xia},
  \citenamefont {Chen}, \citenamefont {Pei}, \citenamefont {Ying},
  \citenamefont {Qi}, \citenamefont {Guo}, \citenamefont {Yang},\ and\
  \citenamefont {Li}}]{Kong2024}%
  \BibitemOpen
  \bibfield  {author} {\bibinfo {author} {\bibfnamefont {X.}~\bibnamefont
  {Kong}}, \bibinfo {author} {\bibfnamefont {Z.}~\bibnamefont {Tao}}, \bibinfo
  {author} {\bibfnamefont {R.}~\bibnamefont {Zhang}}, \bibinfo {author}
  {\bibfnamefont {W.}~\bibnamefont {Xia}}, \bibinfo {author} {\bibfnamefont
  {X.}~\bibnamefont {Chen}}, \bibinfo {author} {\bibfnamefont {C.}~\bibnamefont
  {Pei}}, \bibinfo {author} {\bibfnamefont {T.}~\bibnamefont {Ying}}, \bibinfo
  {author} {\bibfnamefont {Y.}~\bibnamefont {Qi}}, \bibinfo {author}
  {\bibfnamefont {Y.}~\bibnamefont {Guo}}, \bibinfo {author} {\bibfnamefont
  {X.}~\bibnamefont {Yang}},\ and\ \bibinfo {author} {\bibfnamefont
  {S.}~\bibnamefont {Li}},\ }\bibfield  {journal} {\bibinfo  {journal} {CHINESE
  PHYSICS LETTERS}\ }\textbf {\bibinfo {volume} {41}},\ \href
  {https://doi.org/10.1088/0256-307X/41/4/047503}
  {10.1088/0256-307X/41/4/047503} (\bibinfo {year} {2024})\BibitemShut
  {NoStop}%
\bibitem [{\citenamefont {Chen}\ \emph {et~al.}(2024)\citenamefont {Chen},
  \citenamefont {Xing}, \citenamefont {Tan}, \citenamefont {Huang},
  \citenamefont {Zheng}, \citenamefont {Huang}, \citenamefont {Han},
  \citenamefont {Hu}, \citenamefont {Ye}, \citenamefont {Li}, \citenamefont
  {Xiao}, \citenamefont {Lei}, \citenamefont {Qiu}, \citenamefont {Liu},
  \citenamefont {Yang}, \citenamefont {Wang}, \citenamefont {Yan},\ and\
  \citenamefont {Gao}}]{Chen2024}%
  \BibitemOpen
  \bibfield  {author} {\bibinfo {author} {\bibfnamefont {H.}~\bibnamefont
  {Chen}}, \bibinfo {author} {\bibfnamefont {Y.}~\bibnamefont {Xing}}, \bibinfo
  {author} {\bibfnamefont {H.}~\bibnamefont {Tan}}, \bibinfo {author}
  {\bibfnamefont {L.}~\bibnamefont {Huang}}, \bibinfo {author} {\bibfnamefont
  {Q.}~\bibnamefont {Zheng}}, \bibinfo {author} {\bibfnamefont
  {Z.}~\bibnamefont {Huang}}, \bibinfo {author} {\bibfnamefont
  {X.}~\bibnamefont {Han}}, \bibinfo {author} {\bibfnamefont {B.}~\bibnamefont
  {Hu}}, \bibinfo {author} {\bibfnamefont {Y.}~\bibnamefont {Ye}}, \bibinfo
  {author} {\bibfnamefont {Y.}~\bibnamefont {Li}}, \bibinfo {author}
  {\bibfnamefont {Y.}~\bibnamefont {Xiao}}, \bibinfo {author} {\bibfnamefont
  {H.}~\bibnamefont {Lei}}, \bibinfo {author} {\bibfnamefont {X.}~\bibnamefont
  {Qiu}}, \bibinfo {author} {\bibfnamefont {E.}~\bibnamefont {Liu}}, \bibinfo
  {author} {\bibfnamefont {H.}~\bibnamefont {Yang}}, \bibinfo {author}
  {\bibfnamefont {Z.}~\bibnamefont {Wang}}, \bibinfo {author} {\bibfnamefont
  {B.}~\bibnamefont {Yan}},\ and\ \bibinfo {author} {\bibfnamefont {H.-J.}\
  \bibnamefont {Gao}},\ }\bibfield  {journal} {\bibinfo  {journal} {NATURE
  COMMUNICATIONS}\ }\textbf {\bibinfo {volume} {15}},\ \href
  {https://doi.org/10.1038/s41467-024-46729-3} {10.1038/s41467-024-46729-3}
  (\bibinfo {year} {2024})\BibitemShut {NoStop}%
\bibitem [{\citenamefont {Ji}\ \emph {et~al.}(2023)\citenamefont {Ji},
  \citenamefont {Ding}, \citenamefont {Zhan}, \citenamefont {Xiao},
  \citenamefont {Fan}, \citenamefont {Ning},\ and\ \citenamefont
  {Wang}}]{Ji2023}%
  \BibitemOpen
  \bibfield  {author} {\bibinfo {author} {\bibfnamefont {R.}~\bibnamefont
  {Ji}}, \bibinfo {author} {\bibfnamefont {X.}~\bibnamefont {Ding}}, \bibinfo
  {author} {\bibfnamefont {F.}~\bibnamefont {Zhan}}, \bibinfo {author}
  {\bibfnamefont {X.}~\bibnamefont {Xiao}}, \bibinfo {author} {\bibfnamefont
  {J.}~\bibnamefont {Fan}}, \bibinfo {author} {\bibfnamefont {Z.}~\bibnamefont
  {Ning}},\ and\ \bibinfo {author} {\bibfnamefont {R.}~\bibnamefont {Wang}},\
  }\href {https://doi.org/10.1103/PhysRevB.108.205139} {\bibfield  {journal}
  {\bibinfo  {journal} {Phys. Rev. B}\ }\textbf {\bibinfo {volume} {108}},\
  \bibinfo {pages} {205139} (\bibinfo {year} {2023})}\BibitemShut {NoStop}%
\bibitem [{\citenamefont {Mazzola}\ \emph {et~al.}(2023)\citenamefont
  {Mazzola}, \citenamefont {Enzner}, \citenamefont {Eck}, \citenamefont {Bigi},
  \citenamefont {Jugovac}, \citenamefont {Cojocariu}, \citenamefont {Feyer},
  \citenamefont {Shu}, \citenamefont {Pierantozzi}, \citenamefont {De~Vita},
  \citenamefont {Carrara}, \citenamefont {Fujii}, \citenamefont {King},
  \citenamefont {Vinai}, \citenamefont {Orgiani}, \citenamefont {Cacho},
  \citenamefont {Watson}, \citenamefont {Rossi}, \citenamefont {Vobornik},
  \citenamefont {Kong}, \citenamefont {Di~Sante}, \citenamefont {Sangiovanni},\
  and\ \citenamefont {Panaccione}}]{Mazzola2023}%
  \BibitemOpen
  \bibfield  {author} {\bibinfo {author} {\bibfnamefont {F.}~\bibnamefont
  {Mazzola}}, \bibinfo {author} {\bibfnamefont {S.}~\bibnamefont {Enzner}},
  \bibinfo {author} {\bibfnamefont {P.}~\bibnamefont {Eck}}, \bibinfo {author}
  {\bibfnamefont {C.}~\bibnamefont {Bigi}}, \bibinfo {author} {\bibfnamefont
  {M.}~\bibnamefont {Jugovac}}, \bibinfo {author} {\bibfnamefont
  {I.}~\bibnamefont {Cojocariu}}, \bibinfo {author} {\bibfnamefont
  {V.}~\bibnamefont {Feyer}}, \bibinfo {author} {\bibfnamefont
  {Z.}~\bibnamefont {Shu}}, \bibinfo {author} {\bibfnamefont {G.~M.}\
  \bibnamefont {Pierantozzi}}, \bibinfo {author} {\bibfnamefont
  {A.}~\bibnamefont {De~Vita}}, \bibinfo {author} {\bibfnamefont
  {P.}~\bibnamefont {Carrara}}, \bibinfo {author} {\bibfnamefont
  {J.}~\bibnamefont {Fujii}}, \bibinfo {author} {\bibfnamefont {P.~D.~C.}\
  \bibnamefont {King}}, \bibinfo {author} {\bibfnamefont {G.}~\bibnamefont
  {Vinai}}, \bibinfo {author} {\bibfnamefont {P.}~\bibnamefont {Orgiani}},
  \bibinfo {author} {\bibfnamefont {C.}~\bibnamefont {Cacho}}, \bibinfo
  {author} {\bibfnamefont {M.~D.}\ \bibnamefont {Watson}}, \bibinfo {author}
  {\bibfnamefont {G.}~\bibnamefont {Rossi}}, \bibinfo {author} {\bibfnamefont
  {I.}~\bibnamefont {Vobornik}}, \bibinfo {author} {\bibfnamefont
  {T.}~\bibnamefont {Kong}}, \bibinfo {author} {\bibfnamefont {D.}~\bibnamefont
  {Di~Sante}}, \bibinfo {author} {\bibfnamefont {G.}~\bibnamefont
  {Sangiovanni}},\ and\ \bibinfo {author} {\bibfnamefont {G.}~\bibnamefont
  {Panaccione}},\ }\href {https://doi.org/10.1021/acs.nanolett.3c02022}
  {\bibfield  {journal} {\bibinfo  {journal} {NANO LETTERS}\ }\textbf {\bibinfo
  {volume} {23}},\ \bibinfo {pages} {8035} (\bibinfo {year}
  {2023})}\BibitemShut {NoStop}%
\bibitem [{\citenamefont {Huang}\ \emph {et~al.}(2023)\citenamefont {Huang},
  \citenamefont {Kong}, \citenamefont {Zheng}, \citenamefont {Xing},
  \citenamefont {Chen}, \citenamefont {Li}, \citenamefont {Hu}, \citenamefont
  {Zhu}, \citenamefont {Qiao}, \citenamefont {Zhang}, \citenamefont {Cheng},
  \citenamefont {Cheng}, \citenamefont {Qiu}, \citenamefont {Liu},
  \citenamefont {Lei}, \citenamefont {Lin}, \citenamefont {Wang}, \citenamefont
  {Yang}, \citenamefont {Ji},\ and\ \citenamefont {Gao}}]{Huang2023}%
  \BibitemOpen
  \bibfield  {author} {\bibinfo {author} {\bibfnamefont {L.}~\bibnamefont
  {Huang}}, \bibinfo {author} {\bibfnamefont {X.}~\bibnamefont {Kong}},
  \bibinfo {author} {\bibfnamefont {Q.}~\bibnamefont {Zheng}}, \bibinfo
  {author} {\bibfnamefont {Y.}~\bibnamefont {Xing}}, \bibinfo {author}
  {\bibfnamefont {H.}~\bibnamefont {Chen}}, \bibinfo {author} {\bibfnamefont
  {Y.}~\bibnamefont {Li}}, \bibinfo {author} {\bibfnamefont {Z.}~\bibnamefont
  {Hu}}, \bibinfo {author} {\bibfnamefont {S.}~\bibnamefont {Zhu}}, \bibinfo
  {author} {\bibfnamefont {J.}~\bibnamefont {Qiao}}, \bibinfo {author}
  {\bibfnamefont {Y.-Y.}\ \bibnamefont {Zhang}}, \bibinfo {author}
  {\bibfnamefont {H.}~\bibnamefont {Cheng}}, \bibinfo {author} {\bibfnamefont
  {Z.}~\bibnamefont {Cheng}}, \bibinfo {author} {\bibfnamefont
  {X.}~\bibnamefont {Qiu}}, \bibinfo {author} {\bibfnamefont {E.}~\bibnamefont
  {Liu}}, \bibinfo {author} {\bibfnamefont {H.}~\bibnamefont {Lei}}, \bibinfo
  {author} {\bibfnamefont {X.}~\bibnamefont {Lin}}, \bibinfo {author}
  {\bibfnamefont {Z.}~\bibnamefont {Wang}}, \bibinfo {author} {\bibfnamefont
  {H.}~\bibnamefont {Yang}}, \bibinfo {author} {\bibfnamefont {W.}~\bibnamefont
  {Ji}},\ and\ \bibinfo {author} {\bibfnamefont {H.-J.}\ \bibnamefont {Gao}},\
  }\bibfield  {journal} {\bibinfo  {journal} {NATURE COMMUNICATIONS}\ }\textbf
  {\bibinfo {volume} {14}},\ \href {https://doi.org/10.1038/s41467-023-40942-2}
  {10.1038/s41467-023-40942-2} (\bibinfo {year} {2023})\BibitemShut {NoStop}%
\bibitem [{\citenamefont {Ovalle}\ \emph {et~al.}(2022)\citenamefont {Ovalle},
  \citenamefont {Pezo},\ and\ \citenamefont {Manchon}}]{Ovalle2022}%
  \BibitemOpen
  \bibfield  {author} {\bibinfo {author} {\bibfnamefont {D.~G.}\ \bibnamefont
  {Ovalle}}, \bibinfo {author} {\bibfnamefont {A.}~\bibnamefont {Pezo}},\ and\
  \bibinfo {author} {\bibfnamefont {A.}~\bibnamefont {Manchon}},\ }\href
  {https://doi.org/10.1103/PhysRevB.106.214435} {\bibfield  {journal} {\bibinfo
   {journal} {Phys. Rev. B}\ }\textbf {\bibinfo {volume} {106}},\ \bibinfo
  {pages} {214435} (\bibinfo {year} {2022})}\BibitemShut {NoStop}%
\bibitem [{\citenamefont {Hu}\ \emph {et~al.}(2022)\citenamefont {Hu},
  \citenamefont {Wu}, \citenamefont {Yang}, \citenamefont {Gao}, \citenamefont
  {Plumb}, \citenamefont {Schnyder}, \citenamefont {Xie}, \citenamefont {Ma},\
  and\ \citenamefont {Shi}}]{Hu2022}%
  \BibitemOpen
  \bibfield  {author} {\bibinfo {author} {\bibfnamefont {Y.}~\bibnamefont
  {Hu}}, \bibinfo {author} {\bibfnamefont {X.}~\bibnamefont {Wu}}, \bibinfo
  {author} {\bibfnamefont {Y.}~\bibnamefont {Yang}}, \bibinfo {author}
  {\bibfnamefont {S.}~\bibnamefont {Gao}}, \bibinfo {author} {\bibfnamefont
  {N.~C.}\ \bibnamefont {Plumb}}, \bibinfo {author} {\bibfnamefont {A.~P.}\
  \bibnamefont {Schnyder}}, \bibinfo {author} {\bibfnamefont {W.}~\bibnamefont
  {Xie}}, \bibinfo {author} {\bibfnamefont {J.}~\bibnamefont {Ma}},\ and\
  \bibinfo {author} {\bibfnamefont {M.}~\bibnamefont {Shi}},\ }\bibfield
  {journal} {\bibinfo  {journal} {SCIENCE ADVANCES}\ }\textbf {\bibinfo
  {volume} {8}},\ \href {https://doi.org/10.1126/sciadv.add2024}
  {10.1126/sciadv.add2024} (\bibinfo {year} {2022})\BibitemShut {NoStop}%
\bibitem [{\citenamefont {Chen}\ \emph {et~al.}(2022)\citenamefont {Chen},
  \citenamefont {Hanke}, \citenamefont {Hoffmann}, \citenamefont {Bihlmayer},
  \citenamefont {Mokrousov}, \citenamefont {Bluegel}, \citenamefont
  {Schneider},\ and\ \citenamefont {Tusche}}]{Chen2022}%
  \BibitemOpen
  \bibfield  {author} {\bibinfo {author} {\bibfnamefont {Y.-J.}\ \bibnamefont
  {Chen}}, \bibinfo {author} {\bibfnamefont {J.-P.}\ \bibnamefont {Hanke}},
  \bibinfo {author} {\bibfnamefont {M.}~\bibnamefont {Hoffmann}}, \bibinfo
  {author} {\bibfnamefont {G.}~\bibnamefont {Bihlmayer}}, \bibinfo {author}
  {\bibfnamefont {Y.}~\bibnamefont {Mokrousov}}, \bibinfo {author}
  {\bibfnamefont {S.}~\bibnamefont {Bluegel}}, \bibinfo {author} {\bibfnamefont
  {C.~M.}\ \bibnamefont {Schneider}},\ and\ \bibinfo {author} {\bibfnamefont
  {C.}~\bibnamefont {Tusche}},\ }\bibfield  {journal} {\bibinfo  {journal}
  {NATURE COMMUNICATIONS}\ }\textbf {\bibinfo {volume} {13}},\ \href
  {https://doi.org/10.1038/s41467-022-32948-z} {10.1038/s41467-022-32948-z}
  (\bibinfo {year} {2022})\BibitemShut {NoStop}%
\bibitem [{\citenamefont {Buccheri}\ \emph {et~al.}(2022)\citenamefont
  {Buccheri}, \citenamefont {Egger},\ and\ \citenamefont
  {De~Martino}}]{Buccheri2022}%
  \BibitemOpen
  \bibfield  {author} {\bibinfo {author} {\bibfnamefont {F.}~\bibnamefont
  {Buccheri}}, \bibinfo {author} {\bibfnamefont {R.}~\bibnamefont {Egger}},\
  and\ \bibinfo {author} {\bibfnamefont {A.}~\bibnamefont {De~Martino}},\
  }\href {https://doi.org/10.1103/PhysRevB.106.045413} {\bibfield  {journal}
  {\bibinfo  {journal} {Phys. Rev. B}\ }\textbf {\bibinfo {volume} {106}},\
  \bibinfo {pages} {045413} (\bibinfo {year} {2022})}\BibitemShut {NoStop}%
\bibitem [{\citenamefont {Zheng}\ \emph {et~al.}(2021)\citenamefont {Zheng},
  \citenamefont {Chen},\ and\ \citenamefont {Xing}}]{Zheng2021}%
  \BibitemOpen
  \bibfield  {author} {\bibinfo {author} {\bibfnamefont {Y.}~\bibnamefont
  {Zheng}}, \bibinfo {author} {\bibfnamefont {W.}~\bibnamefont {Chen}},\ and\
  \bibinfo {author} {\bibfnamefont {D.~Y.}\ \bibnamefont {Xing}},\ }\href
  {https://doi.org/10.1103/PhysRevB.104.075420} {\bibfield  {journal} {\bibinfo
   {journal} {Phys. Rev. B}\ }\textbf {\bibinfo {volume} {104}},\ \bibinfo
  {pages} {075420} (\bibinfo {year} {2021})}\BibitemShut {NoStop}%
\bibitem [{\citenamefont {Wang}(2021)}]{Wang2021a}%
  \BibitemOpen
  \bibfield  {author} {\bibinfo {author} {\bibfnamefont {L.-L.}\ \bibnamefont
  {Wang}},\ }\href {https://doi.org/10.1103/PhysRevB.103.075105} {\bibfield
  {journal} {\bibinfo  {journal} {Phys. Rev. B}\ }\textbf {\bibinfo {volume}
  {103}},\ \bibinfo {pages} {075105} (\bibinfo {year} {2021})}\BibitemShut
  {NoStop}%
\bibitem [{\citenamefont {Murthy}\ \emph {et~al.}(2020)\citenamefont {Murthy},
  \citenamefont {Fertig},\ and\ \citenamefont {Shimshoni}}]{Murthy2020}%
  \BibitemOpen
  \bibfield  {author} {\bibinfo {author} {\bibfnamefont {G.}~\bibnamefont
  {Murthy}}, \bibinfo {author} {\bibfnamefont {H.~A.}\ \bibnamefont {Fertig}},\
  and\ \bibinfo {author} {\bibfnamefont {E.}~\bibnamefont {Shimshoni}},\ }\href
  {https://doi.org/10.1103/PhysRevResearch.2.013367} {\bibfield  {journal}
  {\bibinfo  {journal} {Phys. Rev. Res.}\ }\textbf {\bibinfo {volume} {2}},\
  \bibinfo {pages} {013367} (\bibinfo {year} {2020})}\BibitemShut {NoStop}%
\bibitem [{\citenamefont {Shvetsov}\ \emph {et~al.}(2020)\citenamefont
  {Shvetsov}, \citenamefont {Esin}, \citenamefont {Barash}, \citenamefont
  {Timonina}, \citenamefont {Kolesnikov},\ and\ \citenamefont
  {Deviatov}}]{Shvetsov2020}%
  \BibitemOpen
  \bibfield  {author} {\bibinfo {author} {\bibfnamefont {O.~O.}\ \bibnamefont
  {Shvetsov}}, \bibinfo {author} {\bibfnamefont {V.~D.}\ \bibnamefont {Esin}},
  \bibinfo {author} {\bibfnamefont {Y.~S.}\ \bibnamefont {Barash}}, \bibinfo
  {author} {\bibfnamefont {A.~V.}\ \bibnamefont {Timonina}}, \bibinfo {author}
  {\bibfnamefont {N.~N.}\ \bibnamefont {Kolesnikov}},\ and\ \bibinfo {author}
  {\bibfnamefont {E.~V.}\ \bibnamefont {Deviatov}},\ }\href
  {https://doi.org/10.1103/PhysRevB.101.035304} {\bibfield  {journal} {\bibinfo
   {journal} {Phys. Rev. B}\ }\textbf {\bibinfo {volume} {101}},\ \bibinfo
  {pages} {035304} (\bibinfo {year} {2020})}\BibitemShut {NoStop}%
\bibitem [{\citenamefont {Yang}\ \emph {et~al.}(2019)\citenamefont {Yang},
  \citenamefont {Yang}, \citenamefont {Liu}, \citenamefont {Sun}, \citenamefont
  {Chen}, \citenamefont {Peng}, \citenamefont {Schmidt}, \citenamefont
  {Prabhakaran}, \citenamefont {Bernevig}, \citenamefont {Felser},
  \citenamefont {Yan},\ and\ \citenamefont {Chen}}]{Yang2019}%
  \BibitemOpen
  \bibfield  {author} {\bibinfo {author} {\bibfnamefont {H.~F.}\ \bibnamefont
  {Yang}}, \bibinfo {author} {\bibfnamefont {L.~X.}\ \bibnamefont {Yang}},
  \bibinfo {author} {\bibfnamefont {Z.~K.}\ \bibnamefont {Liu}}, \bibinfo
  {author} {\bibfnamefont {Y.}~\bibnamefont {Sun}}, \bibinfo {author}
  {\bibfnamefont {C.}~\bibnamefont {Chen}}, \bibinfo {author} {\bibfnamefont
  {H.}~\bibnamefont {Peng}}, \bibinfo {author} {\bibfnamefont {M.}~\bibnamefont
  {Schmidt}}, \bibinfo {author} {\bibfnamefont {D.}~\bibnamefont
  {Prabhakaran}}, \bibinfo {author} {\bibfnamefont {B.~A.}\ \bibnamefont
  {Bernevig}}, \bibinfo {author} {\bibfnamefont {C.}~\bibnamefont {Felser}},
  \bibinfo {author} {\bibfnamefont {B.~H.}\ \bibnamefont {Yan}},\ and\ \bibinfo
  {author} {\bibfnamefont {Y.~L.}\ \bibnamefont {Chen}},\ }\href
  {https://doi.org/10.1038/s41467-019-11491-4} {\bibfield  {journal} {\bibinfo
  {journal} {Nature Communications}\ }\textbf {\bibinfo {volume} {10}},\
  \bibinfo {pages} {3478} (\bibinfo {year} {2019})}\BibitemShut {NoStop}%
\bibitem [{\citenamefont {Quirk}\ \emph {et~al.}(2023)\citenamefont {Quirk},
  \citenamefont {Cheng}, \citenamefont {Manna}, \citenamefont {Felser},
  \citenamefont {Yao},\ and\ \citenamefont {Ong}}]{Quirk2023}%
  \BibitemOpen
  \bibfield  {author} {\bibinfo {author} {\bibfnamefont {N.~P.}\ \bibnamefont
  {Quirk}}, \bibinfo {author} {\bibfnamefont {G.}~\bibnamefont {Cheng}},
  \bibinfo {author} {\bibfnamefont {K.}~\bibnamefont {Manna}}, \bibinfo
  {author} {\bibfnamefont {C.}~\bibnamefont {Felser}}, \bibinfo {author}
  {\bibfnamefont {N.}~\bibnamefont {Yao}},\ and\ \bibinfo {author}
  {\bibfnamefont {N.~P.}\ \bibnamefont {Ong}},\ }\bibfield  {journal} {\bibinfo
   {journal} {NATURE COMMUNICATIONS}\ }\textbf {\bibinfo {volume} {14}},\ \href
  {https://doi.org/10.1038/s41467-023-42222-5} {10.1038/s41467-023-42222-5}
  (\bibinfo {year} {2023})\BibitemShut {NoStop}%
\bibitem [{\citenamefont {Cheng}\ \emph {et~al.}(2024)\citenamefont {Cheng},
  \citenamefont {Yan}, \citenamefont {Shi}, \citenamefont {Lou}, \citenamefont
  {Fedorov}, \citenamefont {Behnami}, \citenamefont {Yuan}, \citenamefont
  {Yang}, \citenamefont {Wang}, \citenamefont {Cheng}, \citenamefont {Xu},
  \citenamefont {Xu}, \citenamefont {Xia}, \citenamefont {Pavlovskii},
  \citenamefont {Peets}, \citenamefont {Zhao}, \citenamefont {Wan},
  \citenamefont {Burkhardt}, \citenamefont {Guo}, \citenamefont {Li},
  \citenamefont {Felser}, \citenamefont {Yang},\ and\ \citenamefont
  {Buechner}}]{Cheng2024}%
  \BibitemOpen
  \bibfield  {author} {\bibinfo {author} {\bibfnamefont {E.}~\bibnamefont
  {Cheng}}, \bibinfo {author} {\bibfnamefont {L.}~\bibnamefont {Yan}}, \bibinfo
  {author} {\bibfnamefont {X.}~\bibnamefont {Shi}}, \bibinfo {author}
  {\bibfnamefont {R.}~\bibnamefont {Lou}}, \bibinfo {author} {\bibfnamefont
  {A.}~\bibnamefont {Fedorov}}, \bibinfo {author} {\bibfnamefont
  {M.}~\bibnamefont {Behnami}}, \bibinfo {author} {\bibfnamefont
  {J.}~\bibnamefont {Yuan}}, \bibinfo {author} {\bibfnamefont {P.}~\bibnamefont
  {Yang}}, \bibinfo {author} {\bibfnamefont {B.}~\bibnamefont {Wang}}, \bibinfo
  {author} {\bibfnamefont {J.-G.}\ \bibnamefont {Cheng}}, \bibinfo {author}
  {\bibfnamefont {Y.}~\bibnamefont {Xu}}, \bibinfo {author} {\bibfnamefont
  {Y.}~\bibnamefont {Xu}}, \bibinfo {author} {\bibfnamefont {W.}~\bibnamefont
  {Xia}}, \bibinfo {author} {\bibfnamefont {N.}~\bibnamefont {Pavlovskii}},
  \bibinfo {author} {\bibfnamefont {D.~C.}\ \bibnamefont {Peets}}, \bibinfo
  {author} {\bibfnamefont {W.}~\bibnamefont {Zhao}}, \bibinfo {author}
  {\bibfnamefont {Y.}~\bibnamefont {Wan}}, \bibinfo {author} {\bibfnamefont
  {U.}~\bibnamefont {Burkhardt}}, \bibinfo {author} {\bibfnamefont
  {Y.}~\bibnamefont {Guo}}, \bibinfo {author} {\bibfnamefont {S.}~\bibnamefont
  {Li}}, \bibinfo {author} {\bibfnamefont {C.}~\bibnamefont {Felser}}, \bibinfo
  {author} {\bibfnamefont {W.}~\bibnamefont {Yang}},\ and\ \bibinfo {author}
  {\bibfnamefont {B.}~\bibnamefont {Buechner}},\ }\bibfield  {journal}
  {\bibinfo  {journal} {NATURE COMMUNICATIONS}\ }\textbf {\bibinfo {volume}
  {15}},\ \href {https://doi.org/10.1038/s41467-024-45658-5}
  {10.1038/s41467-024-45658-5} (\bibinfo {year} {2024})\BibitemShut {NoStop}%
\bibitem [{\citenamefont {Wu}\ \emph {et~al.}(2023)\citenamefont {Wu},
  \citenamefont {Chi}, \citenamefont {Zuo}, \citenamefont {Xu}, \citenamefont
  {Zhao}, \citenamefont {Luo},\ and\ \citenamefont {Zhu}}]{Wu2023}%
  \BibitemOpen
  \bibfield  {author} {\bibinfo {author} {\bibfnamefont {L.}~\bibnamefont
  {Wu}}, \bibinfo {author} {\bibfnamefont {S.}~\bibnamefont {Chi}}, \bibinfo
  {author} {\bibfnamefont {H.}~\bibnamefont {Zuo}}, \bibinfo {author}
  {\bibfnamefont {G.}~\bibnamefont {Xu}}, \bibinfo {author} {\bibfnamefont
  {L.}~\bibnamefont {Zhao}}, \bibinfo {author} {\bibfnamefont {Y.}~\bibnamefont
  {Luo}},\ and\ \bibinfo {author} {\bibfnamefont {Z.}~\bibnamefont {Zhu}},\
  }\bibfield  {journal} {\bibinfo  {journal} {NPJ QUANTUM MATERIALS}\ }\textbf
  {\bibinfo {volume} {8}},\ \href {https://doi.org/10.1038/s41535-023-00537-y}
  {10.1038/s41535-023-00537-y} (\bibinfo {year} {2023})\BibitemShut {NoStop}%
\bibitem [{\citenamefont {Wang}\ \emph
  {et~al.}(2022{\natexlab{a}})\citenamefont {Wang}, \citenamefont {Huang},
  \citenamefont {Hsu}, \citenamefont {Namiki}, \citenamefont {Chang},
  \citenamefont {Chuang}, \citenamefont {Lin}, \citenamefont {Sasagawa},
  \citenamefont {Madhavan},\ and\ \citenamefont {Okada}}]{Wang2022}%
  \BibitemOpen
  \bibfield  {author} {\bibinfo {author} {\bibfnamefont {Z.}~\bibnamefont
  {Wang}}, \bibinfo {author} {\bibfnamefont {C.-Y.}\ \bibnamefont {Huang}},
  \bibinfo {author} {\bibfnamefont {C.-H.}\ \bibnamefont {Hsu}}, \bibinfo
  {author} {\bibfnamefont {H.}~\bibnamefont {Namiki}}, \bibinfo {author}
  {\bibfnamefont {T.-R.}\ \bibnamefont {Chang}}, \bibinfo {author}
  {\bibfnamefont {F.-C.}\ \bibnamefont {Chuang}}, \bibinfo {author}
  {\bibfnamefont {H.}~\bibnamefont {Lin}}, \bibinfo {author} {\bibfnamefont
  {T.}~\bibnamefont {Sasagawa}}, \bibinfo {author} {\bibfnamefont
  {V.}~\bibnamefont {Madhavan}},\ and\ \bibinfo {author} {\bibfnamefont
  {Y.}~\bibnamefont {Okada}},\ }\href
  {https://doi.org/10.1103/PhysRevB.105.075110} {\bibfield  {journal} {\bibinfo
   {journal} {Phys. Rev. B}\ }\textbf {\bibinfo {volume} {105}},\ \bibinfo
  {pages} {075110} (\bibinfo {year} {2022}{\natexlab{a}})}\BibitemShut
  {NoStop}%
\bibitem [{\citenamefont {Wadge}\ \emph {et~al.}(2022)\citenamefont {Wadge},
  \citenamefont {Kowalski}, \citenamefont {Autieri}, \citenamefont {Iwanowski},
  \citenamefont {Hruban}, \citenamefont {Olszowska}, \citenamefont {Rosmus},
  \citenamefont {Ko\l{}odziej},\ and\ \citenamefont {Wi\ifmmode~\acute{s}\else
  \'{s}\fi{}niewski}}]{Wadge2022}%
  \BibitemOpen
  \bibfield  {author} {\bibinfo {author} {\bibfnamefont {A.~S.}\ \bibnamefont
  {Wadge}}, \bibinfo {author} {\bibfnamefont {B.~J.}\ \bibnamefont {Kowalski}},
  \bibinfo {author} {\bibfnamefont {C.}~\bibnamefont {Autieri}}, \bibinfo
  {author} {\bibfnamefont {P.}~\bibnamefont {Iwanowski}}, \bibinfo {author}
  {\bibfnamefont {A.}~\bibnamefont {Hruban}}, \bibinfo {author} {\bibfnamefont
  {N.}~\bibnamefont {Olszowska}}, \bibinfo {author} {\bibfnamefont
  {M.}~\bibnamefont {Rosmus}}, \bibinfo {author} {\bibfnamefont
  {J.}~\bibnamefont {Ko\l{}odziej}},\ and\ \bibinfo {author} {\bibfnamefont
  {A.}~\bibnamefont {Wi\ifmmode~\acute{s}\else \'{s}\fi{}niewski}},\ }\href
  {https://doi.org/10.1103/PhysRevB.105.235304} {\bibfield  {journal} {\bibinfo
   {journal} {Phys. Rev. B}\ }\textbf {\bibinfo {volume} {105}},\ \bibinfo
  {pages} {235304} (\bibinfo {year} {2022})}\BibitemShut {NoStop}%
\bibitem [{\citenamefont {Xu}\ \emph {et~al.}(2015)\citenamefont {Xu},
  \citenamefont {Belopolski}, \citenamefont {Alidoust}, \citenamefont
  {Neupane}, \citenamefont {Bian}, \citenamefont {Zhang}, \citenamefont
  {Sankar}, \citenamefont {Chang}, \citenamefont {Yuan}, \citenamefont {Lee},
  \citenamefont {Huang}, \citenamefont {Zheng}, \citenamefont {Ma},
  \citenamefont {Sanchez}, \citenamefont {Wang}, \citenamefont {Bansil},
  \citenamefont {Chou}, \citenamefont {Shibayev}, \citenamefont {Lin},
  \citenamefont {Jia},\ and\ \citenamefont {Hasan}}]{Xu2015a}%
  \BibitemOpen
  \bibfield  {author} {\bibinfo {author} {\bibfnamefont {S.-Y.}\ \bibnamefont
  {Xu}}, \bibinfo {author} {\bibfnamefont {I.}~\bibnamefont {Belopolski}},
  \bibinfo {author} {\bibfnamefont {N.}~\bibnamefont {Alidoust}}, \bibinfo
  {author} {\bibfnamefont {M.}~\bibnamefont {Neupane}}, \bibinfo {author}
  {\bibfnamefont {G.}~\bibnamefont {Bian}}, \bibinfo {author} {\bibfnamefont
  {C.}~\bibnamefont {Zhang}}, \bibinfo {author} {\bibfnamefont
  {R.}~\bibnamefont {Sankar}}, \bibinfo {author} {\bibfnamefont
  {G.}~\bibnamefont {Chang}}, \bibinfo {author} {\bibfnamefont
  {Z.}~\bibnamefont {Yuan}}, \bibinfo {author} {\bibfnamefont {C.-C.}\
  \bibnamefont {Lee}}, \bibinfo {author} {\bibfnamefont {S.-M.}\ \bibnamefont
  {Huang}}, \bibinfo {author} {\bibfnamefont {H.}~\bibnamefont {Zheng}},
  \bibinfo {author} {\bibfnamefont {J.}~\bibnamefont {Ma}}, \bibinfo {author}
  {\bibfnamefont {D.~S.}\ \bibnamefont {Sanchez}}, \bibinfo {author}
  {\bibfnamefont {B.}~\bibnamefont {Wang}}, \bibinfo {author} {\bibfnamefont
  {A.}~\bibnamefont {Bansil}}, \bibinfo {author} {\bibfnamefont
  {F.}~\bibnamefont {Chou}}, \bibinfo {author} {\bibfnamefont {P.~P.}\
  \bibnamefont {Shibayev}}, \bibinfo {author} {\bibfnamefont {H.}~\bibnamefont
  {Lin}}, \bibinfo {author} {\bibfnamefont {S.}~\bibnamefont {Jia}},\ and\
  \bibinfo {author} {\bibfnamefont {M.~Z.}\ \bibnamefont {Hasan}},\ }\href
  {https://doi.org/10.1126/science.aaa9297} {\bibfield  {journal} {\bibinfo
  {journal} {SCIENCE}\ }\textbf {\bibinfo {volume} {349}},\ \bibinfo {pages}
  {613} (\bibinfo {year} {2015})}\BibitemShut {NoStop}%
\bibitem [{\citenamefont {Lv}\ \emph {et~al.}(2015)\citenamefont {Lv},
  \citenamefont {Weng}, \citenamefont {Fu}, \citenamefont {Wang}, \citenamefont
  {Miao}, \citenamefont {Ma}, \citenamefont {Richard}, \citenamefont {Huang},
  \citenamefont {Zhao}, \citenamefont {Chen}, \citenamefont {Fang},
  \citenamefont {Dai}, \citenamefont {Qian},\ and\ \citenamefont
  {Ding}}]{Lv2015a}%
  \BibitemOpen
  \bibfield  {author} {\bibinfo {author} {\bibfnamefont {B.~Q.}\ \bibnamefont
  {Lv}}, \bibinfo {author} {\bibfnamefont {H.~M.}\ \bibnamefont {Weng}},
  \bibinfo {author} {\bibfnamefont {B.~B.}\ \bibnamefont {Fu}}, \bibinfo
  {author} {\bibfnamefont {X.~P.}\ \bibnamefont {Wang}}, \bibinfo {author}
  {\bibfnamefont {H.}~\bibnamefont {Miao}}, \bibinfo {author} {\bibfnamefont
  {J.}~\bibnamefont {Ma}}, \bibinfo {author} {\bibfnamefont {P.}~\bibnamefont
  {Richard}}, \bibinfo {author} {\bibfnamefont {X.~C.}\ \bibnamefont {Huang}},
  \bibinfo {author} {\bibfnamefont {L.~X.}\ \bibnamefont {Zhao}}, \bibinfo
  {author} {\bibfnamefont {G.~F.}\ \bibnamefont {Chen}}, \bibinfo {author}
  {\bibfnamefont {Z.}~\bibnamefont {Fang}}, \bibinfo {author} {\bibfnamefont
  {X.}~\bibnamefont {Dai}}, \bibinfo {author} {\bibfnamefont {T.}~\bibnamefont
  {Qian}},\ and\ \bibinfo {author} {\bibfnamefont {H.}~\bibnamefont {Ding}},\
  }\href {https://doi.org/10.1103/PhysRevX.5.031013} {\bibfield  {journal}
  {\bibinfo  {journal} {Phys. Rev. X}\ }\textbf {\bibinfo {volume} {5}},\
  \bibinfo {pages} {031013} (\bibinfo {year} {2015})}\BibitemShut {NoStop}%
\bibitem [{\citenamefont {Lv}\ \emph {et~al.}(2021)\citenamefont {Lv},
  \citenamefont {Qian},\ and\ \citenamefont {Ding}}]{Lv2021}%
  \BibitemOpen
  \bibfield  {author} {\bibinfo {author} {\bibfnamefont {B.}~\bibnamefont
  {Lv}}, \bibinfo {author} {\bibfnamefont {T.}~\bibnamefont {Qian}},\ and\
  \bibinfo {author} {\bibfnamefont {H.}~\bibnamefont {Ding}},\ }\href
  {https://doi.org/10.1103/revmodphys.93.025002} {\bibfield  {journal}
  {\bibinfo  {journal} {Reviews of Modern Physics}\ }\textbf {\bibinfo {volume}
  {93}},\ \bibinfo {pages} {025002} (\bibinfo {year} {2021})}\BibitemShut
  {NoStop}%
\bibitem [{\citenamefont {Bernevig}\ \emph {et~al.}(2022)\citenamefont
  {Bernevig}, \citenamefont {Felser},\ and\ \citenamefont
  {Beidenkopf}}]{Bernevig2022}%
  \BibitemOpen
  \bibfield  {author} {\bibinfo {author} {\bibfnamefont {B.~A.}\ \bibnamefont
  {Bernevig}}, \bibinfo {author} {\bibfnamefont {C.}~\bibnamefont {Felser}},\
  and\ \bibinfo {author} {\bibfnamefont {H.}~\bibnamefont {Beidenkopf}},\
  }\href {https://doi.org/10.1038/s41586-021-04105-x} {\bibfield  {journal}
  {\bibinfo  {journal} {Nature}\ }\textbf {\bibinfo {volume} {603}},\ \bibinfo
  {pages} {41} (\bibinfo {year} {2022})}\BibitemShut {NoStop}%
\bibitem [{\citenamefont {da~Silva~Neto}(2019)}]{SilvaNeto2019}%
  \BibitemOpen
  \bibfield  {author} {\bibinfo {author} {\bibfnamefont {E.~H.}\ \bibnamefont
  {da~Silva~Neto}},\ }\href {https://doi.org/10.1126/science.aax6190}
  {\bibfield  {journal} {\bibinfo  {journal} {Science}\ }\textbf {\bibinfo
  {volume} {365}},\ \bibinfo {pages} {1248} (\bibinfo {year}
  {2019})}\BibitemShut {NoStop}%
\bibitem [{\citenamefont {Moll}\ \emph {et~al.}(2016)\citenamefont {Moll},
  \citenamefont {Nair}, \citenamefont {Helm}, \citenamefont {Potter},
  \citenamefont {Kimchi}, \citenamefont {Vishwanath},\ and\ \citenamefont
  {Analytis}}]{Moll2016}%
  \BibitemOpen
  \bibfield  {author} {\bibinfo {author} {\bibfnamefont {P.~J.~W.}\
  \bibnamefont {Moll}}, \bibinfo {author} {\bibfnamefont {N.~L.}\ \bibnamefont
  {Nair}}, \bibinfo {author} {\bibfnamefont {T.}~\bibnamefont {Helm}}, \bibinfo
  {author} {\bibfnamefont {A.~C.}\ \bibnamefont {Potter}}, \bibinfo {author}
  {\bibfnamefont {I.}~\bibnamefont {Kimchi}}, \bibinfo {author} {\bibfnamefont
  {A.}~\bibnamefont {Vishwanath}},\ and\ \bibinfo {author} {\bibfnamefont
  {J.~G.}\ \bibnamefont {Analytis}},\ }\href
  {https://doi.org/10.1038/nature18276} {\bibfield  {journal} {\bibinfo
  {journal} {Nature}\ }\textbf {\bibinfo {volume} {535}},\ \bibinfo {pages}
  {266} (\bibinfo {year} {2016})}\BibitemShut {NoStop}%
\bibitem [{\citenamefont {Huang}\ \emph {et~al.}(2016)\citenamefont {Huang},
  \citenamefont {Xu}, \citenamefont {Belopolski}, \citenamefont {Lee},
  \citenamefont {Chang}, \citenamefont {Chang}, \citenamefont {Wang},
  \citenamefont {Alidoust}, \citenamefont {Bian}, \citenamefont {Neupane},
  \citenamefont {Sanchez}, \citenamefont {Zheng}, \citenamefont {Jeng},
  \citenamefont {Bansil}, \citenamefont {Neupert}, \citenamefont {Lin},\ and\
  \citenamefont {Hasan}}]{Huang2016}%
  \BibitemOpen
  \bibfield  {author} {\bibinfo {author} {\bibfnamefont {S.-M.}\ \bibnamefont
  {Huang}}, \bibinfo {author} {\bibfnamefont {S.-Y.}\ \bibnamefont {Xu}},
  \bibinfo {author} {\bibfnamefont {I.}~\bibnamefont {Belopolski}}, \bibinfo
  {author} {\bibfnamefont {C.-C.}\ \bibnamefont {Lee}}, \bibinfo {author}
  {\bibfnamefont {G.}~\bibnamefont {Chang}}, \bibinfo {author} {\bibfnamefont
  {T.-R.}\ \bibnamefont {Chang}}, \bibinfo {author} {\bibfnamefont
  {B.}~\bibnamefont {Wang}}, \bibinfo {author} {\bibfnamefont {N.}~\bibnamefont
  {Alidoust}}, \bibinfo {author} {\bibfnamefont {G.}~\bibnamefont {Bian}},
  \bibinfo {author} {\bibfnamefont {M.}~\bibnamefont {Neupane}}, \bibinfo
  {author} {\bibfnamefont {D.}~\bibnamefont {Sanchez}}, \bibinfo {author}
  {\bibfnamefont {H.}~\bibnamefont {Zheng}}, \bibinfo {author} {\bibfnamefont
  {H.-T.}\ \bibnamefont {Jeng}}, \bibinfo {author} {\bibfnamefont
  {A.}~\bibnamefont {Bansil}}, \bibinfo {author} {\bibfnamefont
  {T.}~\bibnamefont {Neupert}}, \bibinfo {author} {\bibfnamefont
  {H.}~\bibnamefont {Lin}},\ and\ \bibinfo {author} {\bibfnamefont {M.~Z.}\
  \bibnamefont {Hasan}},\ }\href {https://doi.org/10.1073/pnas.1514581113}
  {\bibfield  {journal} {\bibinfo  {journal} {Proceedings of the National
  Academy of Sciences}\ }\textbf {\bibinfo {volume} {113}},\ \bibinfo {pages}
  {1180} (\bibinfo {year} {2016})}\BibitemShut {NoStop}%
\bibitem [{\citenamefont {Chen}\ \emph {et~al.}(2020)\citenamefont {Chen},
  \citenamefont {Zilberberg},\ and\ \citenamefont {Chen}}]{Chen2020}%
  \BibitemOpen
  \bibfield  {author} {\bibinfo {author} {\bibfnamefont {G.}~\bibnamefont
  {Chen}}, \bibinfo {author} {\bibfnamefont {O.}~\bibnamefont {Zilberberg}},\
  and\ \bibinfo {author} {\bibfnamefont {W.}~\bibnamefont {Chen}},\ }\href
  {https://doi.org/10.1103/PhysRevB.101.125407} {\bibfield  {journal} {\bibinfo
   {journal} {Phys. Rev. B}\ }\textbf {\bibinfo {volume} {101}},\ \bibinfo
  {pages} {125407} (\bibinfo {year} {2020})}\BibitemShut {NoStop}%
\bibitem [{\citenamefont {Miyazaki}\ \emph {et~al.}(2022)\citenamefont
  {Miyazaki}, \citenamefont {Yokouchi}, \citenamefont {Shibata}, \citenamefont
  {Chen}, \citenamefont {Arisawa}, \citenamefont {Mizoguchi}, \citenamefont
  {Saitoh},\ and\ \citenamefont {Shiomi}}]{Miyazaki2022}%
  \BibitemOpen
  \bibfield  {author} {\bibinfo {author} {\bibfnamefont {Y.}~\bibnamefont
  {Miyazaki}}, \bibinfo {author} {\bibfnamefont {T.}~\bibnamefont {Yokouchi}},
  \bibinfo {author} {\bibfnamefont {K.}~\bibnamefont {Shibata}}, \bibinfo
  {author} {\bibfnamefont {Y.}~\bibnamefont {Chen}}, \bibinfo {author}
  {\bibfnamefont {H.}~\bibnamefont {Arisawa}}, \bibinfo {author} {\bibfnamefont
  {T.}~\bibnamefont {Mizoguchi}}, \bibinfo {author} {\bibfnamefont
  {E.}~\bibnamefont {Saitoh}},\ and\ \bibinfo {author} {\bibfnamefont
  {Y.}~\bibnamefont {Shiomi}},\ }\href
  {https://doi.org/10.1103/PhysRevResearch.4.L022002} {\bibfield  {journal}
  {\bibinfo  {journal} {Phys. Rev. Res.}\ }\textbf {\bibinfo {volume} {4}},\
  \bibinfo {pages} {L022002} (\bibinfo {year} {2022})}\BibitemShut {NoStop}%
\bibitem [{\citenamefont {Li}\ \emph {et~al.}(2020)\citenamefont {Li},
  \citenamefont {Wang}, \citenamefont {Li}, \citenamefont {Zheng},
  \citenamefont {Brinkman}, \citenamefont {Yu},\ and\ \citenamefont
  {Liao}}]{Li2020}%
  \BibitemOpen
  \bibfield  {author} {\bibinfo {author} {\bibfnamefont {C.-Z.}\ \bibnamefont
  {Li}}, \bibinfo {author} {\bibfnamefont {A.-Q.}\ \bibnamefont {Wang}},
  \bibinfo {author} {\bibfnamefont {C.}~\bibnamefont {Li}}, \bibinfo {author}
  {\bibfnamefont {W.-Z.}\ \bibnamefont {Zheng}}, \bibinfo {author}
  {\bibfnamefont {A.}~\bibnamefont {Brinkman}}, \bibinfo {author}
  {\bibfnamefont {D.-P.}\ \bibnamefont {Yu}},\ and\ \bibinfo {author}
  {\bibfnamefont {Z.-M.}\ \bibnamefont {Liao}},\ }\bibfield  {journal}
  {\bibinfo  {journal} {Nature Communications}\ }\textbf {\bibinfo {volume}
  {11}},\ \href {https://doi.org/10.1038/s41467-020-15010-8}
  {10.1038/s41467-020-15010-8} (\bibinfo {year} {2020})\BibitemShut {NoStop}%
\bibitem [{\citenamefont {Tokura}\ and\ \citenamefont
  {Nagaosa}(2018)}]{Tokura2018}%
  \BibitemOpen
  \bibfield  {author} {\bibinfo {author} {\bibfnamefont {Y.}~\bibnamefont
  {Tokura}}\ and\ \bibinfo {author} {\bibfnamefont {N.}~\bibnamefont
  {Nagaosa}},\ }\href {https://doi.org/10.1038/s41467-018-05759-4} {\bibfield
  {journal} {\bibinfo  {journal} {Nature Communications}\ }\textbf {\bibinfo
  {volume} {9}},\ \bibinfo {pages} {3740} (\bibinfo {year} {2018})}\BibitemShut
  {NoStop}%
\bibitem [{\citenamefont {Ideue}\ \emph {et~al.}(2017)\citenamefont {Ideue},
  \citenamefont {Hamamoto}, \citenamefont {Koshikawa}, \citenamefont {Ezawa},
  \citenamefont {Shimizu}, \citenamefont {Kaneko}, \citenamefont {Tokura},
  \citenamefont {Nagaosa},\ and\ \citenamefont {Iwasa}}]{Ideue2017}%
  \BibitemOpen
  \bibfield  {author} {\bibinfo {author} {\bibfnamefont {T.}~\bibnamefont
  {Ideue}}, \bibinfo {author} {\bibfnamefont {K.}~\bibnamefont {Hamamoto}},
  \bibinfo {author} {\bibfnamefont {S.}~\bibnamefont {Koshikawa}}, \bibinfo
  {author} {\bibfnamefont {M.}~\bibnamefont {Ezawa}}, \bibinfo {author}
  {\bibfnamefont {S.}~\bibnamefont {Shimizu}}, \bibinfo {author} {\bibfnamefont
  {Y.}~\bibnamefont {Kaneko}}, \bibinfo {author} {\bibfnamefont
  {Y.}~\bibnamefont {Tokura}}, \bibinfo {author} {\bibfnamefont
  {N.}~\bibnamefont {Nagaosa}},\ and\ \bibinfo {author} {\bibfnamefont
  {Y.}~\bibnamefont {Iwasa}},\ }\href {https://doi.org/10.1038/nphys4056}
  {\bibfield  {journal} {\bibinfo  {journal} {Nature Physics}\ }\textbf
  {\bibinfo {volume} {13}},\ \bibinfo {pages} {578} (\bibinfo {year}
  {2017})}\BibitemShut {NoStop}%
\bibitem [{\citenamefont {Itahashi}\ \emph {et~al.}(2020)\citenamefont
  {Itahashi}, \citenamefont {Ideue}, \citenamefont {Saito}, \citenamefont
  {Shimizu}, \citenamefont {Ouchi}, \citenamefont {Nojima},\ and\ \citenamefont
  {Iwasa}}]{Itahashi2020}%
  \BibitemOpen
  \bibfield  {author} {\bibinfo {author} {\bibfnamefont {Y.~M.}\ \bibnamefont
  {Itahashi}}, \bibinfo {author} {\bibfnamefont {T.}~\bibnamefont {Ideue}},
  \bibinfo {author} {\bibfnamefont {Y.}~\bibnamefont {Saito}}, \bibinfo
  {author} {\bibfnamefont {S.}~\bibnamefont {Shimizu}}, \bibinfo {author}
  {\bibfnamefont {T.}~\bibnamefont {Ouchi}}, \bibinfo {author} {\bibfnamefont
  {T.}~\bibnamefont {Nojima}},\ and\ \bibinfo {author} {\bibfnamefont
  {Y.}~\bibnamefont {Iwasa}},\ }\href {https://doi.org/10.1126/sciadv.aay9120}
  {\bibfield  {journal} {\bibinfo  {journal} {Science Advances}\ }\textbf
  {\bibinfo {volume} {6}},\ \bibinfo {pages} {eaay9120} (\bibinfo {year}
  {2020})}\BibitemShut {NoStop}%
\bibitem [{\citenamefont {Li}\ \emph {et~al.}(2021)\citenamefont {Li},
  \citenamefont {Li}, \citenamefont {Li}, \citenamefont {Fang}, \citenamefont
  {Yang}, \citenamefont {Wen}, \citenamefont {Zheng}, \citenamefont {Zhang},
  \citenamefont {He}, \citenamefont {Manchon}, \citenamefont {Cheng},\ and\
  \citenamefont {Zhang}}]{Li2021a}%
  \BibitemOpen
  \bibfield  {author} {\bibinfo {author} {\bibfnamefont {Y.}~\bibnamefont
  {Li}}, \bibinfo {author} {\bibfnamefont {Y.}~\bibnamefont {Li}}, \bibinfo
  {author} {\bibfnamefont {P.}~\bibnamefont {Li}}, \bibinfo {author}
  {\bibfnamefont {B.}~\bibnamefont {Fang}}, \bibinfo {author} {\bibfnamefont
  {X.}~\bibnamefont {Yang}}, \bibinfo {author} {\bibfnamefont {Y.}~\bibnamefont
  {Wen}}, \bibinfo {author} {\bibfnamefont {D.-x.}\ \bibnamefont {Zheng}},
  \bibinfo {author} {\bibfnamefont {C.-h.}\ \bibnamefont {Zhang}}, \bibinfo
  {author} {\bibfnamefont {X.}~\bibnamefont {He}}, \bibinfo {author}
  {\bibfnamefont {A.}~\bibnamefont {Manchon}}, \bibinfo {author} {\bibfnamefont
  {Z.-H.}\ \bibnamefont {Cheng}},\ and\ \bibinfo {author} {\bibfnamefont
  {X.-x.}\ \bibnamefont {Zhang}},\ }\href
  {https://doi.org/10.1038/s41467-020-20840-7} {\bibfield  {journal} {\bibinfo
  {journal} {Nature Communications}\ }\textbf {\bibinfo {volume} {12}},\
  \bibinfo {pages} {540} (\bibinfo {year} {2021})}\BibitemShut {NoStop}%
\bibitem [{\citenamefont {Yasuda}\ \emph {et~al.}(2019)\citenamefont {Yasuda},
  \citenamefont {Yasuda}, \citenamefont {Liang}, \citenamefont {Yoshimi},
  \citenamefont {Tsukazaki}, \citenamefont {Takahashi}, \citenamefont
  {Nagaosa}, \citenamefont {Kawasaki},\ and\ \citenamefont
  {Tokura}}]{Yasuda2019}%
  \BibitemOpen
  \bibfield  {author} {\bibinfo {author} {\bibfnamefont {K.}~\bibnamefont
  {Yasuda}}, \bibinfo {author} {\bibfnamefont {H.}~\bibnamefont {Yasuda}},
  \bibinfo {author} {\bibfnamefont {T.}~\bibnamefont {Liang}}, \bibinfo
  {author} {\bibfnamefont {R.}~\bibnamefont {Yoshimi}}, \bibinfo {author}
  {\bibfnamefont {A.}~\bibnamefont {Tsukazaki}}, \bibinfo {author}
  {\bibfnamefont {K.~S.}\ \bibnamefont {Takahashi}}, \bibinfo {author}
  {\bibfnamefont {N.}~\bibnamefont {Nagaosa}}, \bibinfo {author} {\bibfnamefont
  {M.}~\bibnamefont {Kawasaki}},\ and\ \bibinfo {author} {\bibfnamefont
  {Y.}~\bibnamefont {Tokura}},\ }\href
  {https://doi.org/10.1038/s41467-019-10658-3} {\bibfield  {journal} {\bibinfo
  {journal} {Nature Communications}\ }\textbf {\bibinfo {volume} {10}},\
  \bibinfo {pages} {2734} (\bibinfo {year} {2019})}\BibitemShut {NoStop}%
\bibitem [{\citenamefont {Choe}\ \emph {et~al.}(2019)\citenamefont {Choe},
  \citenamefont {Jin}, \citenamefont {Kim}, \citenamefont {Choi}, \citenamefont
  {Jo}, \citenamefont {Oh}, \citenamefont {Park}, \citenamefont {Jin},
  \citenamefont {Koo}, \citenamefont {Min}, \citenamefont {Hong}, \citenamefont
  {Lee}, \citenamefont {Baek},\ and\ \citenamefont {Yoo}}]{Choe2019}%
  \BibitemOpen
  \bibfield  {author} {\bibinfo {author} {\bibfnamefont {D.}~\bibnamefont
  {Choe}}, \bibinfo {author} {\bibfnamefont {M.-J.}\ \bibnamefont {Jin}},
  \bibinfo {author} {\bibfnamefont {S.-I.}\ \bibnamefont {Kim}}, \bibinfo
  {author} {\bibfnamefont {H.-J.}\ \bibnamefont {Choi}}, \bibinfo {author}
  {\bibfnamefont {J.}~\bibnamefont {Jo}}, \bibinfo {author} {\bibfnamefont
  {I.}~\bibnamefont {Oh}}, \bibinfo {author} {\bibfnamefont {J.}~\bibnamefont
  {Park}}, \bibinfo {author} {\bibfnamefont {H.}~\bibnamefont {Jin}}, \bibinfo
  {author} {\bibfnamefont {H.~C.}\ \bibnamefont {Koo}}, \bibinfo {author}
  {\bibfnamefont {B.-C.}\ \bibnamefont {Min}}, \bibinfo {author} {\bibfnamefont
  {S.}~\bibnamefont {Hong}}, \bibinfo {author} {\bibfnamefont {H.-W.}\
  \bibnamefont {Lee}}, \bibinfo {author} {\bibfnamefont {S.-H.}\ \bibnamefont
  {Baek}},\ and\ \bibinfo {author} {\bibfnamefont {J.-W.}\ \bibnamefont
  {Yoo}},\ }\href {https://doi.org/10.1038/s41467-019-12466-1} {\bibfield
  {journal} {\bibinfo  {journal} {Nature Communications}\ }\textbf {\bibinfo
  {volume} {10}},\ \bibinfo {pages} {4510} (\bibinfo {year}
  {2019})}\BibitemShut {NoStop}%
\bibitem [{\citenamefont {Shim}\ \emph {et~al.}(2022)\citenamefont {Shim},
  \citenamefont {Mehraeen}, \citenamefont {Sklenar}, \citenamefont {Oh},
  \citenamefont {Gibbons}, \citenamefont {Saglam}, \citenamefont {Hoffmann},
  \citenamefont {Zhang},\ and\ \citenamefont {Mason}}]{Shim2022}%
  \BibitemOpen
  \bibfield  {author} {\bibinfo {author} {\bibfnamefont {S.}~\bibnamefont
  {Shim}}, \bibinfo {author} {\bibfnamefont {M.}~\bibnamefont {Mehraeen}},
  \bibinfo {author} {\bibfnamefont {J.}~\bibnamefont {Sklenar}}, \bibinfo
  {author} {\bibfnamefont {J.}~\bibnamefont {Oh}}, \bibinfo {author}
  {\bibfnamefont {J.}~\bibnamefont {Gibbons}}, \bibinfo {author} {\bibfnamefont
  {H.}~\bibnamefont {Saglam}}, \bibinfo {author} {\bibfnamefont
  {A.}~\bibnamefont {Hoffmann}}, \bibinfo {author} {\bibfnamefont {S.~S.-L.}\
  \bibnamefont {Zhang}},\ and\ \bibinfo {author} {\bibfnamefont
  {N.}~\bibnamefont {Mason}},\ }\href
  {https://doi.org/10.1103/PhysRevX.12.021069} {\bibfield  {journal} {\bibinfo
  {journal} {Phys. Rev. X}\ }\textbf {\bibinfo {volume} {12}},\ \bibinfo
  {pages} {021069} (\bibinfo {year} {2022})}\BibitemShut {NoStop}%
\bibitem [{\citenamefont {Ye}\ \emph {et~al.}(2022)\citenamefont {Ye},
  \citenamefont {Xie}, \citenamefont {Lv}, \citenamefont {Huang}, \citenamefont
  {Yang}, \citenamefont {Jiang}, \citenamefont {Liu}, \citenamefont {Zhu},
  \citenamefont {Qiu}, \citenamefont {Tong}, \citenamefont {Zhou},
  \citenamefont {Hsu}, \citenamefont {Chang}, \citenamefont {Lin},
  \citenamefont {Li}, \citenamefont {Yang}, \citenamefont {Wang}, \citenamefont
  {Jiang},\ and\ \citenamefont {Renshaw~Wang}}]{Ye2022}%
  \BibitemOpen
  \bibfield  {author} {\bibinfo {author} {\bibfnamefont {C.}~\bibnamefont
  {Ye}}, \bibinfo {author} {\bibfnamefont {X.}~\bibnamefont {Xie}}, \bibinfo
  {author} {\bibfnamefont {W.}~\bibnamefont {Lv}}, \bibinfo {author}
  {\bibfnamefont {K.}~\bibnamefont {Huang}}, \bibinfo {author} {\bibfnamefont
  {A.~J.}\ \bibnamefont {Yang}}, \bibinfo {author} {\bibfnamefont
  {S.}~\bibnamefont {Jiang}}, \bibinfo {author} {\bibfnamefont
  {X.}~\bibnamefont {Liu}}, \bibinfo {author} {\bibfnamefont {D.}~\bibnamefont
  {Zhu}}, \bibinfo {author} {\bibfnamefont {X.}~\bibnamefont {Qiu}}, \bibinfo
  {author} {\bibfnamefont {M.}~\bibnamefont {Tong}}, \bibinfo {author}
  {\bibfnamefont {T.}~\bibnamefont {Zhou}}, \bibinfo {author} {\bibfnamefont
  {C.-H.}\ \bibnamefont {Hsu}}, \bibinfo {author} {\bibfnamefont
  {G.}~\bibnamefont {Chang}}, \bibinfo {author} {\bibfnamefont
  {H.}~\bibnamefont {Lin}}, \bibinfo {author} {\bibfnamefont {P.}~\bibnamefont
  {Li}}, \bibinfo {author} {\bibfnamefont {K.}~\bibnamefont {Yang}}, \bibinfo
  {author} {\bibfnamefont {Z.}~\bibnamefont {Wang}}, \bibinfo {author}
  {\bibfnamefont {T.}~\bibnamefont {Jiang}},\ and\ \bibinfo {author}
  {\bibfnamefont {X.}~\bibnamefont {Renshaw~Wang}},\ }\href
  {https://doi.org/10.1021/acs.nanolett.1c04756} {\bibfield  {journal}
  {\bibinfo  {journal} {Nano Letters}\ }\textbf {\bibinfo {volume} {22}},\
  \bibinfo {pages} {1366} (\bibinfo {year} {2022})}\BibitemShut {NoStop}%
\bibitem [{\citenamefont {Yasuda}\ \emph {et~al.}(2020)\citenamefont {Yasuda},
  \citenamefont {Morimoto}, \citenamefont {Yoshimi}, \citenamefont {Mogi},
  \citenamefont {Tsukazaki}, \citenamefont {Kawamura}, \citenamefont
  {Takahashi}, \citenamefont {Kawasaki}, \citenamefont {Nagaosa},\ and\
  \citenamefont {Tokura}}]{Yasuda2020}%
  \BibitemOpen
  \bibfield  {author} {\bibinfo {author} {\bibfnamefont {K.}~\bibnamefont
  {Yasuda}}, \bibinfo {author} {\bibfnamefont {T.}~\bibnamefont {Morimoto}},
  \bibinfo {author} {\bibfnamefont {R.}~\bibnamefont {Yoshimi}}, \bibinfo
  {author} {\bibfnamefont {M.}~\bibnamefont {Mogi}}, \bibinfo {author}
  {\bibfnamefont {A.}~\bibnamefont {Tsukazaki}}, \bibinfo {author}
  {\bibfnamefont {M.}~\bibnamefont {Kawamura}}, \bibinfo {author}
  {\bibfnamefont {K.~S.}\ \bibnamefont {Takahashi}}, \bibinfo {author}
  {\bibfnamefont {M.}~\bibnamefont {Kawasaki}}, \bibinfo {author}
  {\bibfnamefont {N.}~\bibnamefont {Nagaosa}},\ and\ \bibinfo {author}
  {\bibfnamefont {Y.}~\bibnamefont {Tokura}},\ }\href
  {https://doi.org/10.1038/s41565-020-0733-2} {\bibfield  {journal} {\bibinfo
  {journal} {Nature Nanotechnology}\ }\textbf {\bibinfo {volume} {15}},\
  \bibinfo {pages} {831} (\bibinfo {year} {2020})}\BibitemShut {NoStop}%
\bibitem [{\citenamefont {Zhang}\ \emph {et~al.}(2022)\citenamefont {Zhang},
  \citenamefont {Wang}, \citenamefont {Cao}, \citenamefont {Wang},
  \citenamefont {Zhou}, \citenamefont {Watanabe}, \citenamefont {Taniguchi},
  \citenamefont {Yan},\ and\ \citenamefont {bo~Gao}}]{Zhang2022}%
  \BibitemOpen
  \bibfield  {author} {\bibinfo {author} {\bibfnamefont {Z.}~\bibnamefont
  {Zhang}}, \bibinfo {author} {\bibfnamefont {N.}~\bibnamefont {Wang}},
  \bibinfo {author} {\bibfnamefont {N.}~\bibnamefont {Cao}}, \bibinfo {author}
  {\bibfnamefont {A.}~\bibnamefont {Wang}}, \bibinfo {author} {\bibfnamefont
  {X.}~\bibnamefont {Zhou}}, \bibinfo {author} {\bibfnamefont {K.}~\bibnamefont
  {Watanabe}}, \bibinfo {author} {\bibfnamefont {T.}~\bibnamefont {Taniguchi}},
  \bibinfo {author} {\bibfnamefont {B.}~\bibnamefont {Yan}},\ and\ \bibinfo
  {author} {\bibfnamefont {W.}~\bibnamefont {bo~Gao}},\ }\bibfield  {journal}
  {\bibinfo  {journal} {Nature Communications}\ }\textbf {\bibinfo {volume}
  {13}},\ \href {https://doi.org/10.1038/s41467-022-33705-y}
  {10.1038/s41467-022-33705-y} (\bibinfo {year} {2022})\BibitemShut {NoStop}%
\bibitem [{\citenamefont {Wang}\ \emph {et~al.}(2020)\citenamefont {Wang},
  \citenamefont {Fu}, \citenamefont {Zhang}, \citenamefont {Zhao},\ and\
  \citenamefont {Zhang}}]{Wang2020a}%
  \BibitemOpen
  \bibfield  {author} {\bibinfo {author} {\bibfnamefont {Z.}~\bibnamefont
  {Wang}}, \bibinfo {author} {\bibfnamefont {Z.}~\bibnamefont {Fu}}, \bibinfo
  {author} {\bibfnamefont {P.}~\bibnamefont {Zhang}}, \bibinfo {author}
  {\bibfnamefont {X.-G.}\ \bibnamefont {Zhao}},\ and\ \bibinfo {author}
  {\bibfnamefont {W.}~\bibnamefont {Zhang}},\ }\href
  {https://doi.org/10.1103/PhysRevB.101.245313} {\bibfield  {journal} {\bibinfo
   {journal} {Phys. Rev. B}\ }\textbf {\bibinfo {volume} {101}},\ \bibinfo
  {pages} {245313} (\bibinfo {year} {2020})}\BibitemShut {NoStop}%
\bibitem [{\citenamefont {Zeng}\ \emph {et~al.}(2021)\citenamefont {Zeng},
  \citenamefont {Nandy},\ and\ \citenamefont {Tewari}}]{Zeng2021}%
  \BibitemOpen
  \bibfield  {author} {\bibinfo {author} {\bibfnamefont {C.}~\bibnamefont
  {Zeng}}, \bibinfo {author} {\bibfnamefont {S.}~\bibnamefont {Nandy}},\ and\
  \bibinfo {author} {\bibfnamefont {S.}~\bibnamefont {Tewari}},\ }\href
  {https://doi.org/10.1103/PhysRevB.103.245119} {\bibfield  {journal} {\bibinfo
   {journal} {Phys. Rev. B}\ }\textbf {\bibinfo {volume} {103}},\ \bibinfo
  {pages} {245119} (\bibinfo {year} {2021})}\BibitemShut {NoStop}%
\bibitem [{\citenamefont {Das}\ \emph {et~al.}(2022)\citenamefont {Das},
  \citenamefont {Das},\ and\ \citenamefont {Agarwal}}]{Das2022}%
  \BibitemOpen
  \bibfield  {author} {\bibinfo {author} {\bibfnamefont {S.}~\bibnamefont
  {Das}}, \bibinfo {author} {\bibfnamefont {K.}~\bibnamefont {Das}},\ and\
  \bibinfo {author} {\bibfnamefont {A.}~\bibnamefont {Agarwal}},\ }\href
  {https://doi.org/10.1103/physrevb.105.235408} {\bibfield  {journal} {\bibinfo
   {journal} {Physical Review B}\ }\textbf {\bibinfo {volume} {105}},\ \bibinfo
  {pages} {235408} (\bibinfo {year} {2022})}\BibitemShut {NoStop}%
\bibitem [{\citenamefont {Rikken}\ \emph {et~al.}(2001)\citenamefont {Rikken},
  \citenamefont {Fölling},\ and\ \citenamefont {Wyder}}]{Rikken2001}%
  \BibitemOpen
  \bibfield  {author} {\bibinfo {author} {\bibfnamefont {G.}~\bibnamefont
  {Rikken}}, \bibinfo {author} {\bibfnamefont {J.}~\bibnamefont {Fölling}},\
  and\ \bibinfo {author} {\bibfnamefont {P.}~\bibnamefont {Wyder}},\ }\href
  {https://doi.org/10.1103/PhysRevLett.87.236602} {\bibfield  {journal}
  {\bibinfo  {journal} {Physical Review Letters}\ }\textbf {\bibinfo {volume}
  {87}},\ \bibinfo {pages} {236602} (\bibinfo {year} {2001})}\BibitemShut
  {NoStop}%
\bibitem [{\citenamefont {Krstić}\ \emph {et~al.}(2002)\citenamefont
  {Krstić}, \citenamefont {Roth}, \citenamefont {Burghard}, \citenamefont
  {Kern},\ and\ \citenamefont {Rikken}}]{Krstic2002}%
  \BibitemOpen
  \bibfield  {author} {\bibinfo {author} {\bibfnamefont {V.}~\bibnamefont
  {Krstić}}, \bibinfo {author} {\bibfnamefont {S.}~\bibnamefont {Roth}},
  \bibinfo {author} {\bibfnamefont {M.}~\bibnamefont {Burghard}}, \bibinfo
  {author} {\bibfnamefont {K.}~\bibnamefont {Kern}},\ and\ \bibinfo {author}
  {\bibfnamefont {G.~L. J.~A.}\ \bibnamefont {Rikken}},\ }\href
  {https://doi.org/10.1063/1.1523895} {\bibfield  {journal} {\bibinfo
  {journal} {The Journal of Chemical Physics}\ }\textbf {\bibinfo {volume}
  {117}},\ \bibinfo {pages} {11315} (\bibinfo {year} {2002})}\BibitemShut
  {NoStop}%
\bibitem [{\citenamefont {Wang}\ \emph
  {et~al.}(2022{\natexlab{b}})\citenamefont {Wang}, \citenamefont {Legg},
  \citenamefont {B\"omerich}, \citenamefont {Park}, \citenamefont {Biesenkamp},
  \citenamefont {Taskin}, \citenamefont {Braden}, \citenamefont {Rosch},\ and\
  \citenamefont {Ando}}]{Wang2022a}%
  \BibitemOpen
  \bibfield  {author} {\bibinfo {author} {\bibfnamefont {Y.}~\bibnamefont
  {Wang}}, \bibinfo {author} {\bibfnamefont {H.~F.}\ \bibnamefont {Legg}},
  \bibinfo {author} {\bibfnamefont {T.}~\bibnamefont {B\"omerich}}, \bibinfo
  {author} {\bibfnamefont {J.}~\bibnamefont {Park}}, \bibinfo {author}
  {\bibfnamefont {S.}~\bibnamefont {Biesenkamp}}, \bibinfo {author}
  {\bibfnamefont {A.~A.}\ \bibnamefont {Taskin}}, \bibinfo {author}
  {\bibfnamefont {M.}~\bibnamefont {Braden}}, \bibinfo {author} {\bibfnamefont
  {A.}~\bibnamefont {Rosch}},\ and\ \bibinfo {author} {\bibfnamefont
  {Y.}~\bibnamefont {Ando}},\ }\href
  {https://doi.org/10.1103/PhysRevLett.128.176602} {\bibfield  {journal}
  {\bibinfo  {journal} {Phys. Rev. Lett.}\ }\textbf {\bibinfo {volume} {128}},\
  \bibinfo {pages} {176602} (\bibinfo {year} {2022}{\natexlab{b}})}\BibitemShut
  {NoStop}%
\bibitem [{\citenamefont {Guo}\ \emph {et~al.}(2022)\citenamefont {Guo},
  \citenamefont {Putzke}, \citenamefont {Konyzheva}, \citenamefont {Huang},
  \citenamefont {Gutierrez-Amigo}, \citenamefont {Errea}, \citenamefont {Chen},
  \citenamefont {Vergniory}, \citenamefont {Felser}, \citenamefont {Fischer},
  \citenamefont {Neupert},\ and\ \citenamefont {Moll}}]{Guo2022}%
  \BibitemOpen
  \bibfield  {author} {\bibinfo {author} {\bibfnamefont {C.}~\bibnamefont
  {Guo}}, \bibinfo {author} {\bibfnamefont {C.}~\bibnamefont {Putzke}},
  \bibinfo {author} {\bibfnamefont {S.}~\bibnamefont {Konyzheva}}, \bibinfo
  {author} {\bibfnamefont {X.}~\bibnamefont {Huang}}, \bibinfo {author}
  {\bibfnamefont {M.}~\bibnamefont {Gutierrez-Amigo}}, \bibinfo {author}
  {\bibfnamefont {I.}~\bibnamefont {Errea}}, \bibinfo {author} {\bibfnamefont
  {D.}~\bibnamefont {Chen}}, \bibinfo {author} {\bibfnamefont {M.~G.}\
  \bibnamefont {Vergniory}}, \bibinfo {author} {\bibfnamefont {C.}~\bibnamefont
  {Felser}}, \bibinfo {author} {\bibfnamefont {M.~H.}\ \bibnamefont {Fischer}},
  \bibinfo {author} {\bibfnamefont {T.}~\bibnamefont {Neupert}},\ and\ \bibinfo
  {author} {\bibfnamefont {P.~J.~W.}\ \bibnamefont {Moll}},\ }\href
  {https://doi.org/10.1038/s41586-022-05127-9} {\bibfield  {journal} {\bibinfo
  {journal} {Nature}\ }\textbf {\bibinfo {volume} {611}},\ \bibinfo {pages}
  {461} (\bibinfo {year} {2022})}\BibitemShut {NoStop}%
\bibitem [{\citenamefont {Ma}\ \emph {et~al.}(2018)\citenamefont {Ma},
  \citenamefont {Xu}, \citenamefont {Shen}, \citenamefont {MacNeill},
  \citenamefont {Fatemi}, \citenamefont {Chang}, \citenamefont {Mier~Valdivia},
  \citenamefont {Wu}, \citenamefont {Du}, \citenamefont {Hsu}, \citenamefont
  {Fang}, \citenamefont {Gibson}, \citenamefont {Watanabe}, \citenamefont
  {Taniguchi}, \citenamefont {Cava}, \citenamefont {Kaxiras}, \citenamefont
  {Lu}, \citenamefont {Lin}, \citenamefont {Fu}, \citenamefont {Gedik},\ and\
  \citenamefont {Jarillo-Herrero}}]{Ma2018}%
  \BibitemOpen
  \bibfield  {author} {\bibinfo {author} {\bibfnamefont {Q.}~\bibnamefont
  {Ma}}, \bibinfo {author} {\bibfnamefont {S.-Y.}\ \bibnamefont {Xu}}, \bibinfo
  {author} {\bibfnamefont {H.}~\bibnamefont {Shen}}, \bibinfo {author}
  {\bibfnamefont {D.}~\bibnamefont {MacNeill}}, \bibinfo {author}
  {\bibfnamefont {V.}~\bibnamefont {Fatemi}}, \bibinfo {author} {\bibfnamefont
  {T.-R.}\ \bibnamefont {Chang}}, \bibinfo {author} {\bibfnamefont {A.~M.}\
  \bibnamefont {Mier~Valdivia}}, \bibinfo {author} {\bibfnamefont
  {S.}~\bibnamefont {Wu}}, \bibinfo {author} {\bibfnamefont {Z.}~\bibnamefont
  {Du}}, \bibinfo {author} {\bibfnamefont {C.-H.}\ \bibnamefont {Hsu}},
  \bibinfo {author} {\bibfnamefont {S.}~\bibnamefont {Fang}}, \bibinfo {author}
  {\bibfnamefont {Q.~D.}\ \bibnamefont {Gibson}}, \bibinfo {author}
  {\bibfnamefont {K.}~\bibnamefont {Watanabe}}, \bibinfo {author}
  {\bibfnamefont {T.}~\bibnamefont {Taniguchi}}, \bibinfo {author}
  {\bibfnamefont {R.~J.}\ \bibnamefont {Cava}}, \bibinfo {author}
  {\bibfnamefont {E.}~\bibnamefont {Kaxiras}}, \bibinfo {author} {\bibfnamefont
  {H.-Z.}\ \bibnamefont {Lu}}, \bibinfo {author} {\bibfnamefont
  {H.}~\bibnamefont {Lin}}, \bibinfo {author} {\bibfnamefont {L.}~\bibnamefont
  {Fu}}, \bibinfo {author} {\bibfnamefont {N.}~\bibnamefont {Gedik}},\ and\
  \bibinfo {author} {\bibfnamefont {P.}~\bibnamefont {Jarillo-Herrero}},\
  }\href {https://doi.org/10.1038/s41586-018-0807-6} {\bibfield  {journal}
  {\bibinfo  {journal} {Nature}\ }\textbf {\bibinfo {volume} {565}},\ \bibinfo
  {pages} {337} (\bibinfo {year} {2018})}\BibitemShut {NoStop}%
\bibitem [{\citenamefont {Wang}\ \emph {et~al.}(2023)\citenamefont {Wang},
  \citenamefont {You}, \citenamefont {Wang}, \citenamefont {Zhou},
  \citenamefont {Zhang}, \citenamefont {Lai}, \citenamefont {Feng},
  \citenamefont {Lin}, \citenamefont {Chang},\ and\ \citenamefont
  {Gao}}]{Wang2023a}%
  \BibitemOpen
  \bibfield  {author} {\bibinfo {author} {\bibfnamefont {N.}~\bibnamefont
  {Wang}}, \bibinfo {author} {\bibfnamefont {J.-Y.}\ \bibnamefont {You}},
  \bibinfo {author} {\bibfnamefont {A.}~\bibnamefont {Wang}}, \bibinfo {author}
  {\bibfnamefont {X.}~\bibnamefont {Zhou}}, \bibinfo {author} {\bibfnamefont
  {Z.}~\bibnamefont {Zhang}}, \bibinfo {author} {\bibfnamefont
  {S.}~\bibnamefont {Lai}}, \bibinfo {author} {\bibfnamefont {Y.-P.}\
  \bibnamefont {Feng}}, \bibinfo {author} {\bibfnamefont {H.}~\bibnamefont
  {Lin}}, \bibinfo {author} {\bibfnamefont {G.}~\bibnamefont {Chang}},\ and\
  \bibinfo {author} {\bibfnamefont {W.-b.}\ \bibnamefont {Gao}},\ }\bibfield
  {journal} {\bibinfo  {journal} {National Science Review}\ }\textbf {\bibinfo
  {volume} {11}},\ \href {https://doi.org/10.1093/nsr/nwad103}
  {10.1093/nsr/nwad103} (\bibinfo {year} {2023})\BibitemShut {NoStop}%
\bibitem [{\citenamefont {Yasuda}\ \emph {et~al.}(2016)\citenamefont {Yasuda},
  \citenamefont {Tsukazaki}, \citenamefont {Yoshimi}, \citenamefont
  {Takahashi}, \citenamefont {Kawasaki},\ and\ \citenamefont
  {Tokura}}]{Yasuda2016}%
  \BibitemOpen
  \bibfield  {author} {\bibinfo {author} {\bibfnamefont {K.}~\bibnamefont
  {Yasuda}}, \bibinfo {author} {\bibfnamefont {A.}~\bibnamefont {Tsukazaki}},
  \bibinfo {author} {\bibfnamefont {R.}~\bibnamefont {Yoshimi}}, \bibinfo
  {author} {\bibfnamefont {K.~S.}\ \bibnamefont {Takahashi}}, \bibinfo {author}
  {\bibfnamefont {M.}~\bibnamefont {Kawasaki}},\ and\ \bibinfo {author}
  {\bibfnamefont {Y.}~\bibnamefont {Tokura}},\ }\href
  {https://doi.org/10.1103/PhysRevLett.117.127202} {\bibfield  {journal}
  {\bibinfo  {journal} {Phys. Rev. Lett.}\ }\textbf {\bibinfo {volume} {117}},\
  \bibinfo {pages} {127202} (\bibinfo {year} {2016})}\BibitemShut {NoStop}%
\bibitem [{\citenamefont {Liu}\ \emph {et~al.}(2019)\citenamefont {Liu},
  \citenamefont {Liang}, \citenamefont {Liu}, \citenamefont {Xu}, \citenamefont
  {Li}, \citenamefont {Chen}, \citenamefont {Pei}, \citenamefont {Shi},
  \citenamefont {Mo}, \citenamefont {Dudin}, \citenamefont {Kim}, \citenamefont
  {Cacho}, \citenamefont {Li}, \citenamefont {Sun}, \citenamefont {Yang},
  \citenamefont {Liu}, \citenamefont {Parkin}, \citenamefont {Felser},\ and\
  \citenamefont {Chen}}]{Liu2019a}%
  \BibitemOpen
  \bibfield  {author} {\bibinfo {author} {\bibfnamefont {D.~F.}\ \bibnamefont
  {Liu}}, \bibinfo {author} {\bibfnamefont {A.~J.}\ \bibnamefont {Liang}},
  \bibinfo {author} {\bibfnamefont {E.~K.}\ \bibnamefont {Liu}}, \bibinfo
  {author} {\bibfnamefont {Q.~N.}\ \bibnamefont {Xu}}, \bibinfo {author}
  {\bibfnamefont {Y.~W.}\ \bibnamefont {Li}}, \bibinfo {author} {\bibfnamefont
  {C.}~\bibnamefont {Chen}}, \bibinfo {author} {\bibfnamefont {D.}~\bibnamefont
  {Pei}}, \bibinfo {author} {\bibfnamefont {W.~J.}\ \bibnamefont {Shi}},
  \bibinfo {author} {\bibfnamefont {S.~K.}\ \bibnamefont {Mo}}, \bibinfo
  {author} {\bibfnamefont {P.}~\bibnamefont {Dudin}}, \bibinfo {author}
  {\bibfnamefont {T.}~\bibnamefont {Kim}}, \bibinfo {author} {\bibfnamefont
  {C.}~\bibnamefont {Cacho}}, \bibinfo {author} {\bibfnamefont
  {G.}~\bibnamefont {Li}}, \bibinfo {author} {\bibfnamefont {Y.}~\bibnamefont
  {Sun}}, \bibinfo {author} {\bibfnamefont {L.~X.}\ \bibnamefont {Yang}},
  \bibinfo {author} {\bibfnamefont {Z.~K.}\ \bibnamefont {Liu}}, \bibinfo
  {author} {\bibfnamefont {S.~S.~P.}\ \bibnamefont {Parkin}}, \bibinfo {author}
  {\bibfnamefont {C.}~\bibnamefont {Felser}},\ and\ \bibinfo {author}
  {\bibfnamefont {Y.~L.}\ \bibnamefont {Chen}},\ }\href
  {https://doi.org/10.1126/science.aav2873} {\bibfield  {journal} {\bibinfo
  {journal} {Science}\ }\textbf {\bibinfo {volume} {365}},\ \bibinfo {pages}
  {1282} (\bibinfo {year} {2019})}\BibitemShut {NoStop}%
\bibitem [{\citenamefont {Howard}\ \emph {et~al.}(2021)\citenamefont {Howard},
  \citenamefont {Jiao}, \citenamefont {Wang}, \citenamefont {Morali},
  \citenamefont {Batabyal}, \citenamefont {Kumar-Nag}, \citenamefont {Avraham},
  \citenamefont {Beidenkopf}, \citenamefont {Vir}, \citenamefont {Liu},
  \citenamefont {Shekhar}, \citenamefont {Felser}, \citenamefont {Hughes},\
  and\ \citenamefont {Madhavan}}]{Howard2021}%
  \BibitemOpen
  \bibfield  {author} {\bibinfo {author} {\bibfnamefont {S.}~\bibnamefont
  {Howard}}, \bibinfo {author} {\bibfnamefont {L.}~\bibnamefont {Jiao}},
  \bibinfo {author} {\bibfnamefont {Z.}~\bibnamefont {Wang}}, \bibinfo {author}
  {\bibfnamefont {N.}~\bibnamefont {Morali}}, \bibinfo {author} {\bibfnamefont
  {R.}~\bibnamefont {Batabyal}}, \bibinfo {author} {\bibfnamefont
  {P.}~\bibnamefont {Kumar-Nag}}, \bibinfo {author} {\bibfnamefont
  {N.}~\bibnamefont {Avraham}}, \bibinfo {author} {\bibfnamefont
  {H.}~\bibnamefont {Beidenkopf}}, \bibinfo {author} {\bibfnamefont
  {P.}~\bibnamefont {Vir}}, \bibinfo {author} {\bibfnamefont {E.}~\bibnamefont
  {Liu}}, \bibinfo {author} {\bibfnamefont {C.}~\bibnamefont {Shekhar}},
  \bibinfo {author} {\bibfnamefont {C.}~\bibnamefont {Felser}}, \bibinfo
  {author} {\bibfnamefont {T.}~\bibnamefont {Hughes}},\ and\ \bibinfo {author}
  {\bibfnamefont {V.}~\bibnamefont {Madhavan}},\ }\bibfield  {journal}
  {\bibinfo  {journal} {NATURE COMMUNICATIONS}\ }\textbf {\bibinfo {volume}
  {12}},\ \href {https://doi.org/10.1038/s41467-021-24561-3}
  {10.1038/s41467-021-24561-3} (\bibinfo {year} {2021})\BibitemShut {NoStop}%
\bibitem [{\citenamefont {Liu}\ \emph {et~al.}(2018)\citenamefont {Liu},
  \citenamefont {Sun}, \citenamefont {Kumar}, \citenamefont {Muechler},
  \citenamefont {Sun}, \citenamefont {Jiao}, \citenamefont {Yang},
  \citenamefont {Liu}, \citenamefont {Liang}, \citenamefont {Xu}, \citenamefont
  {Kroder}, \citenamefont {Sü{\ss}}, \citenamefont {Borrmann}, \citenamefont
  {Shekhar}, \citenamefont {Wang}, \citenamefont {Xi}, \citenamefont {Wang},
  \citenamefont {Schnelle}, \citenamefont {Wirth}, \citenamefont {Chen},
  \citenamefont {Goennenwein},\ and\ \citenamefont {Felser}}]{Liu2018}%
  \BibitemOpen
  \bibfield  {author} {\bibinfo {author} {\bibfnamefont {E.}~\bibnamefont
  {Liu}}, \bibinfo {author} {\bibfnamefont {Y.}~\bibnamefont {Sun}}, \bibinfo
  {author} {\bibfnamefont {N.}~\bibnamefont {Kumar}}, \bibinfo {author}
  {\bibfnamefont {L.}~\bibnamefont {Muechler}}, \bibinfo {author}
  {\bibfnamefont {A.}~\bibnamefont {Sun}}, \bibinfo {author} {\bibfnamefont
  {L.}~\bibnamefont {Jiao}}, \bibinfo {author} {\bibfnamefont {S.-Y.}\
  \bibnamefont {Yang}}, \bibinfo {author} {\bibfnamefont {D.}~\bibnamefont
  {Liu}}, \bibinfo {author} {\bibfnamefont {A.}~\bibnamefont {Liang}}, \bibinfo
  {author} {\bibfnamefont {Q.}~\bibnamefont {Xu}}, \bibinfo {author}
  {\bibfnamefont {J.}~\bibnamefont {Kroder}}, \bibinfo {author} {\bibfnamefont
  {V.}~\bibnamefont {Sü{\ss}}}, \bibinfo {author} {\bibfnamefont
  {H.}~\bibnamefont {Borrmann}}, \bibinfo {author} {\bibfnamefont
  {C.}~\bibnamefont {Shekhar}}, \bibinfo {author} {\bibfnamefont
  {Z.}~\bibnamefont {Wang}}, \bibinfo {author} {\bibfnamefont {C.}~\bibnamefont
  {Xi}}, \bibinfo {author} {\bibfnamefont {W.}~\bibnamefont {Wang}}, \bibinfo
  {author} {\bibfnamefont {W.}~\bibnamefont {Schnelle}}, \bibinfo {author}
  {\bibfnamefont {S.}~\bibnamefont {Wirth}}, \bibinfo {author} {\bibfnamefont
  {Y.}~\bibnamefont {Chen}}, \bibinfo {author} {\bibfnamefont {S.~T.~B.}\
  \bibnamefont {Goennenwein}},\ and\ \bibinfo {author} {\bibfnamefont
  {C.}~\bibnamefont {Felser}},\ }\href
  {https://doi.org/10.1038/s41567-018-0234-5} {\bibfield  {journal} {\bibinfo
  {journal} {Nature Physics}\ }\textbf {\bibinfo {volume} {14}},\ \bibinfo
  {pages} {1125} (\bibinfo {year} {2018})}\BibitemShut {NoStop}%
\bibitem [{\citenamefont {Wang}\ \emph
  {et~al.}(2022{\natexlab{c}})\citenamefont {Wang}, \citenamefont {Zeng},
  \citenamefont {Yuan}, \citenamefont {Zeng}, \citenamefont {Gu}, \citenamefont
  {Xu}, \citenamefont {Wang}, \citenamefont {Han}, \citenamefont {Nomura},
  \citenamefont {Wang}, \citenamefont {Liu}, \citenamefont {Hou},\ and\
  \citenamefont {Ye}}]{Wang2022b}%
  \BibitemOpen
  \bibfield  {author} {\bibinfo {author} {\bibfnamefont {Q.}~\bibnamefont
  {Wang}}, \bibinfo {author} {\bibfnamefont {Y.}~\bibnamefont {Zeng}}, \bibinfo
  {author} {\bibfnamefont {K.}~\bibnamefont {Yuan}}, \bibinfo {author}
  {\bibfnamefont {Q.}~\bibnamefont {Zeng}}, \bibinfo {author} {\bibfnamefont
  {P.}~\bibnamefont {Gu}}, \bibinfo {author} {\bibfnamefont {X.}~\bibnamefont
  {Xu}}, \bibinfo {author} {\bibfnamefont {H.}~\bibnamefont {Wang}}, \bibinfo
  {author} {\bibfnamefont {Z.}~\bibnamefont {Han}}, \bibinfo {author}
  {\bibfnamefont {K.}~\bibnamefont {Nomura}}, \bibinfo {author} {\bibfnamefont
  {W.}~\bibnamefont {Wang}}, \bibinfo {author} {\bibfnamefont {E.}~\bibnamefont
  {Liu}}, \bibinfo {author} {\bibfnamefont {Y.}~\bibnamefont {Hou}},\ and\
  \bibinfo {author} {\bibfnamefont {Y.}~\bibnamefont {Ye}},\ }\href
  {https://doi.org/10.1038/s41928-022-00879-8} {\bibfield  {journal} {\bibinfo
  {journal} {Nature Electronics}\ }\textbf {\bibinfo {volume} {6}},\ \bibinfo
  {pages} {119} (\bibinfo {year} {2022}{\natexlab{c}})}\BibitemShut {NoStop}%
\bibitem [{\citenamefont {Yang}\ \emph {et~al.}(2025)\citenamefont {Yang},
  \citenamefont {Shang}, \citenamefont {Liu}, \citenamefont {Wang},
  \citenamefont {Dong}, \citenamefont {Zeng}, \citenamefont {Lyu},
  \citenamefont {Zhang}, \citenamefont {Liu}, \citenamefont {Wang},
  \citenamefont {Wei}, \citenamefont {Wu}, \citenamefont {Parkin},
  \citenamefont {Liu}, \citenamefont {Felser}, \citenamefont {Liu},\ and\
  \citenamefont {Shen}}]{Yang2025}%
  \BibitemOpen
  \bibfield  {author} {\bibinfo {author} {\bibfnamefont {J.}~\bibnamefont
  {Yang}}, \bibinfo {author} {\bibfnamefont {Y.}~\bibnamefont {Shang}},
  \bibinfo {author} {\bibfnamefont {X.}~\bibnamefont {Liu}}, \bibinfo {author}
  {\bibfnamefont {Y.}~\bibnamefont {Wang}}, \bibinfo {author} {\bibfnamefont
  {X.}~\bibnamefont {Dong}}, \bibinfo {author} {\bibfnamefont {Q.}~\bibnamefont
  {Zeng}}, \bibinfo {author} {\bibfnamefont {M.}~\bibnamefont {Lyu}}, \bibinfo
  {author} {\bibfnamefont {S.}~\bibnamefont {Zhang}}, \bibinfo {author}
  {\bibfnamefont {Y.}~\bibnamefont {Liu}}, \bibinfo {author} {\bibfnamefont
  {B.}~\bibnamefont {Wang}}, \bibinfo {author} {\bibfnamefont {H.}~\bibnamefont
  {Wei}}, \bibinfo {author} {\bibfnamefont {Y.}~\bibnamefont {Wu}}, \bibinfo
  {author} {\bibfnamefont {S.}~\bibnamefont {Parkin}}, \bibinfo {author}
  {\bibfnamefont {G.}~\bibnamefont {Liu}}, \bibinfo {author} {\bibfnamefont
  {C.}~\bibnamefont {Felser}}, \bibinfo {author} {\bibfnamefont
  {E.}~\bibnamefont {Liu}},\ and\ \bibinfo {author} {\bibfnamefont
  {B.}~\bibnamefont {Shen}},\ }\href
  {https://doi.org/10.1038/s41928-025-01364-8} {\bibfield  {journal} {\bibinfo
  {journal} {Nature Electronics}\ }\textbf {\bibinfo {volume} {8}},\ \bibinfo
  {pages} {386} (\bibinfo {year} {2025})}\BibitemShut {NoStop}%
\bibitem [{\citenamefont {Ozawa}\ and\ \citenamefont
  {Nomura}(2019)}]{Ozawa2019}%
  \BibitemOpen
  \bibfield  {author} {\bibinfo {author} {\bibfnamefont {A.}~\bibnamefont
  {Ozawa}}\ and\ \bibinfo {author} {\bibfnamefont {K.}~\bibnamefont {Nomura}},\
  }\href {https://doi.org/10.7566/jpsj.88.123703} {\bibfield  {journal}
  {\bibinfo  {journal} {Journal of the Physical Society of Japan}\ }\textbf
  {\bibinfo {volume} {88}},\ \bibinfo {pages} {123703} (\bibinfo {year}
  {2019})}\BibitemShut {NoStop}%
\bibitem [{\citenamefont {Kaplan}\ \emph
  {et~al.}(2024{\natexlab{a}})\citenamefont {Kaplan}, \citenamefont {Holder},\
  and\ \citenamefont {Yan}}]{Kaplan2024a}%
  \BibitemOpen
  \bibfield  {author} {\bibinfo {author} {\bibfnamefont {D.}~\bibnamefont
  {Kaplan}}, \bibinfo {author} {\bibfnamefont {T.}~\bibnamefont {Holder}},\
  and\ \bibinfo {author} {\bibfnamefont {B.}~\bibnamefont {Yan}},\ }\href
  {https://doi.org/10.1103/physrevlett.132.026301} {\bibfield  {journal}
  {\bibinfo  {journal} {Physical Review Letters}\ }\textbf {\bibinfo {volume}
  {132}},\ \bibinfo {pages} {026301} (\bibinfo {year}
  {2024}{\natexlab{a}})}\BibitemShut {NoStop}%
\bibitem [{\citenamefont {Kaplan}\ \emph
  {et~al.}(2024{\natexlab{b}})\citenamefont {Kaplan}, \citenamefont {Holder},\
  and\ \citenamefont {Yan}}]{Kaplan2024}%
  \BibitemOpen
  \bibfield  {author} {\bibinfo {author} {\bibfnamefont {D.}~\bibnamefont
  {Kaplan}}, \bibinfo {author} {\bibfnamefont {T.}~\bibnamefont {Holder}},\
  and\ \bibinfo {author} {\bibfnamefont {B.}~\bibnamefont {Yan}},\ }\href
  {https://doi.org/10.1103/PhysRevLett.132.026301} {\bibfield  {journal}
  {\bibinfo  {journal} {Phys. Rev. Lett.}\ }\textbf {\bibinfo {volume} {132}},\
  \bibinfo {pages} {026301} (\bibinfo {year} {2024}{\natexlab{b}})}\BibitemShut
  {NoStop}%
\bibitem [{\citenamefont {Ikeda}\ \emph {et~al.}(2021)\citenamefont {Ikeda},
  \citenamefont {Fujiwara}, \citenamefont {Shiogai}, \citenamefont {Seki},
  \citenamefont {Nomura}, \citenamefont {Takanashi},\ and\ \citenamefont
  {Tsukazaki}}]{Ikeda2021a}%
  \BibitemOpen
  \bibfield  {author} {\bibinfo {author} {\bibfnamefont {J.}~\bibnamefont
  {Ikeda}}, \bibinfo {author} {\bibfnamefont {K.}~\bibnamefont {Fujiwara}},
  \bibinfo {author} {\bibfnamefont {J.}~\bibnamefont {Shiogai}}, \bibinfo
  {author} {\bibfnamefont {T.}~\bibnamefont {Seki}}, \bibinfo {author}
  {\bibfnamefont {K.}~\bibnamefont {Nomura}}, \bibinfo {author} {\bibfnamefont
  {K.}~\bibnamefont {Takanashi}},\ and\ \bibinfo {author} {\bibfnamefont
  {A.}~\bibnamefont {Tsukazaki}},\ }\bibfield  {journal} {\bibinfo  {journal}
  {COMMUNICATIONS MATERIALS}\ }\textbf {\bibinfo {volume} {2}},\ \href
  {https://doi.org/10.1038/s43246-021-00122-5} {10.1038/s43246-021-00122-5}
  (\bibinfo {year} {2021})\BibitemShut {NoStop}%
\bibitem [{\citenamefont {Belopolski}\ \emph {et~al.}(2019)\citenamefont
  {Belopolski}, \citenamefont {Manna}, \citenamefont {Sanchez}, \citenamefont
  {Chang}, \citenamefont {Ernst}, \citenamefont {Yin}, \citenamefont {Zhang},
  \citenamefont {Cochran}, \citenamefont {Shumiya}, \citenamefont {Zheng},
  \citenamefont {Singh}, \citenamefont {Bian}, \citenamefont {Multer},
  \citenamefont {Litskevich}, \citenamefont {Zhou}, \citenamefont {Huang},
  \citenamefont {Wang}, \citenamefont {Chang}, \citenamefont {Xu},
  \citenamefont {Bansil}, \citenamefont {Felser}, \citenamefont {Lin},\ and\
  \citenamefont {Hasan}}]{Belopolski2019}%
  \BibitemOpen
  \bibfield  {author} {\bibinfo {author} {\bibfnamefont {I.}~\bibnamefont
  {Belopolski}}, \bibinfo {author} {\bibfnamefont {K.}~\bibnamefont {Manna}},
  \bibinfo {author} {\bibfnamefont {D.~S.}\ \bibnamefont {Sanchez}}, \bibinfo
  {author} {\bibfnamefont {G.}~\bibnamefont {Chang}}, \bibinfo {author}
  {\bibfnamefont {B.}~\bibnamefont {Ernst}}, \bibinfo {author} {\bibfnamefont
  {J.}~\bibnamefont {Yin}}, \bibinfo {author} {\bibfnamefont {S.~S.}\
  \bibnamefont {Zhang}}, \bibinfo {author} {\bibfnamefont {T.}~\bibnamefont
  {Cochran}}, \bibinfo {author} {\bibfnamefont {N.}~\bibnamefont {Shumiya}},
  \bibinfo {author} {\bibfnamefont {H.}~\bibnamefont {Zheng}}, \bibinfo
  {author} {\bibfnamefont {B.}~\bibnamefont {Singh}}, \bibinfo {author}
  {\bibfnamefont {G.}~\bibnamefont {Bian}}, \bibinfo {author} {\bibfnamefont
  {D.}~\bibnamefont {Multer}}, \bibinfo {author} {\bibfnamefont
  {M.}~\bibnamefont {Litskevich}}, \bibinfo {author} {\bibfnamefont
  {X.}~\bibnamefont {Zhou}}, \bibinfo {author} {\bibfnamefont {S.-M.}\
  \bibnamefont {Huang}}, \bibinfo {author} {\bibfnamefont {B.}~\bibnamefont
  {Wang}}, \bibinfo {author} {\bibfnamefont {T.-R.}\ \bibnamefont {Chang}},
  \bibinfo {author} {\bibfnamefont {S.-Y.}\ \bibnamefont {Xu}}, \bibinfo
  {author} {\bibfnamefont {A.}~\bibnamefont {Bansil}}, \bibinfo {author}
  {\bibfnamefont {C.}~\bibnamefont {Felser}}, \bibinfo {author} {\bibfnamefont
  {H.}~\bibnamefont {Lin}},\ and\ \bibinfo {author} {\bibfnamefont {M.~Z.}\
  \bibnamefont {Hasan}},\ }\href {https://doi.org/10.1126/science.aav2327}
  {\bibfield  {journal} {\bibinfo  {journal} {Science}\ }\textbf {\bibinfo
  {volume} {365}},\ \bibinfo {pages} {1278} (\bibinfo {year} {2019})},\ \Eprint
  {https://arxiv.org/abs/https://www.science.org/doi/pdf/10.1126/science.aav2327}
  {https://www.science.org/doi/pdf/10.1126/science.aav2327} \BibitemShut
  {NoStop}%
\end{thebibliography}
%

	\appendix
	\section{Derivation Of $\sigma_2$ }
\begin{figure}
     	\centering
     	\includegraphics[width=1.0\columnwidth]{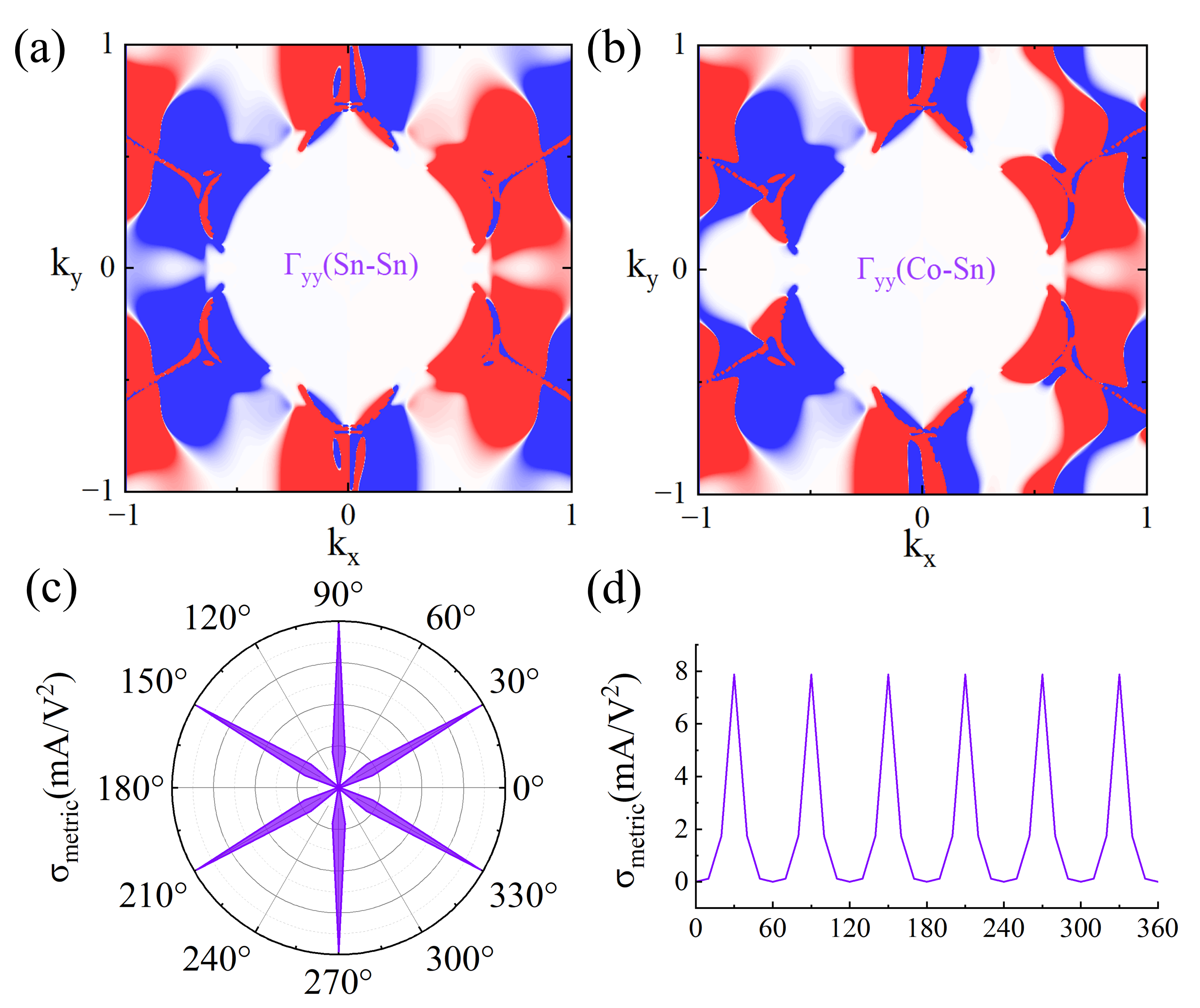}
     	\caption{(a,b) Metric dipole $\Gamma_{yy}=v_yG^{yy}$ for identical (Co-Co) vs. distinct (Co-Sn) surface terminations, The integration of asymmetric $\Gamma_{yy}$ (Co-Sn) gives the non-zero second-order conductivity. (c,d) Second-order conductivity of distinct terminations at zero bias.
     	}\label{fig6}
     \end{figure}

	Here we present the derivation of Eq.~(12).we begin with
	\begin{equation}
	      	j_y = e \sum\int v_y f_2 \, d{\mathbf{k}} + e^2 E^2  \sum\int \frac{\partial G^{yy}_n}{\hbar\partial k_y} f_0 \, d\mathbf{k},
	\end{equation}
     where the first term accounts for the Drude response, and the second captures the contribution from the quantum metric dipole. Substituting $f_2$
     into the Drude term yields
     \begin{equation}
     	\begin{split}
     	&e \sum\int v_y f_2 \, d{\mathbf{k}}=\\
     	&\tau^2e^3E^2\sum_n\int dk_x\int \left(v_y\frac{\partial v_y}{\hbar\partial k_y}\frac{\partial f_0}{\partial \varepsilon}+v_y^3\frac{\partial^2 f_0}{\partial^2 \varepsilon}\right)dk_y.
     	\end{split}
     \end{equation}
      Applying integration by parts to the second term on the right leads to
      \begin{equation}
      	\begin{split}
      		\int v_y^3\frac{\partial^2 f_0}{\partial^2 \varepsilon}dk_y&=\int v_y^2\frac{\partial(\frac{\partial f_0}{\partial \varepsilon})}{\hbar \partial k_y}dk_y\\
      		&=\frac{\partial f_0}{\partial \varepsilon}v_y^2\Big|_{-\infty}^{\infty}-\int 2v_y\frac{\partial v_y}{\hbar\partial k_y}\frac{\partial f_0}{\partial \varepsilon}dk_y
      	\end{split}
      \end{equation}
      Since the first term is identically zero, inserting the second term into Eq.~(A1) and applying integration by parts yields
      \begin{equation}
      	\begin{split}
      	j_{Drude}&=-\tau^2e^3E^2\sum_n\int dk_x\int v_y\frac{\partial v_y}{\hbar\partial k_y}\frac{\partial f_0}{\partial \varepsilon}dk_y\\
      	&=-\tau^2e^3E^2\sum_n\int dk_x \left(f_0 \frac{\partial v_y}{\hbar^2\partial k_y}\Big|_{-\infty}^{\infty}-\int \frac{\partial^2 v_y}{\hbar^2\partial k_y^2} f_0 dk_y\right)\\
      	&=\frac{\tau^2e^3E^2}{\hbar^3}\sum_n\int \frac{\partial^3 \varepsilon_n}{\partial k_y^3} f_0 d\mathbf{k}. 
      	\end{split}
      \end{equation}
     The derivation of the metric dipole contribution follows the semiclassical formalism developed in \cite{Kaplan2024a}. Since the Drude contribution is affected by scattering while the metric contribution is intrinsic, their relative magnitudes are determined by the material-dependent relaxation time: The strength of the Drude response grows with increasing relaxation time($\tau^2$), in contrast to the intrinsic metric contribution. In Fermi-arc-dominated transport, impurity scattering is relatively weak, and here we take the relaxation time as [$10^{-12}$]. As shown in Fig.~\ref{fig6}, once the symmetry between the two surfaces is broken, the integrated metric dipole no longer vanishes and gives rise to a finite second-order conductivity. However, its magnitude remains one to two orders smaller than the Drude contribution discussed in the main text.
     
     \section{Derivation Of $\gamma$ }
     In this section, we derive the expression for the nonreciprocal coefficient $\gamma$ which defined in main text as $R =R_0(1+\gamma I)$. We start from the definition of the second-order current density
     \begin{equation}
     	\begin{split}
     		j=\sigma_1E+\sigma_2E^2.
     	\end{split}
     \end{equation}
    Multiplying both sides of the equation by $L_xL_z$
    allows us to rewrite it in the form of current
    \begin{equation}
   	\begin{split}
   		I&=\sigma_1E+\sigma_2E^2\\
   		&=\frac{1}{R_0}V+\frac{L_xL_z}{L_y^2}\sigma_2V^2,
   		\end{split}
   	\end{equation}
   	with $R_0=\frac{L_y}{L_xL_z\sigma_1}$ and $V=L_yE$. On the other hand, from $R =R_0(1+\gamma I)$ we have
    \begin{equation}
    	V =IR_0+\gamma I^2R_0.
    \end{equation}
   	Substituting B3 into B2 gives 
   	\begin{equation}
   		\begin{split}
   			I&=I+\gamma I^2+\frac{L_xL_z}{L_y^2}\sigma_2R_0^2I^2\\
   				&+\frac{L_xL_z}{L_y^2}\sigma_2(2R_0\gamma I^3+\gamma I^4).
   				\end{split}
   			\end{equation}
     Neglecting the higher-order terms of 
     $I$, we obtain $\gamma=-\frac{1}{L_xL_z}\frac{\sigma_2}{\sigma_1^2}$.
     
     \section{Symmetry constraint on the second-order conductivity tensor under $C_3$ rotation}
     
     We consider a second-order nonlinear current response
     \begin{equation}
     	j_i = \sigma_{ijk} E_j E_k ,
     \end{equation}
     where $\sigma_{ijk}$ is a third-rank conductivity tensor (symmetric in the last two indices, i.e., $\sigma_{ijk}=\sigma_{ikj}$).
    
     In the two-dimensional plane, it is convenient to use the complex basis
     \begin{equation}
     	j_{\pm} = j_x \pm i j_y, \qquad
     	E_{\pm} = E_x \pm i E_y .
     \end{equation}
     Under an in-plane rotation by an angle $\phi$, these quantities transform as
     \begin{equation}
     	j_{\pm} \rightarrow e^{\pm i\phi} j_{\pm}, \qquad 
     	E_{\pm} \rightarrow e^{\pm i\phi} E_{\pm}.
     \end{equation}
     
     The general quadratic form allowed by the tensor symmetry is
     \begin{equation}
     	j_+ = \eta_1 E_+^2 + \eta_2 E_+ E_- + \eta_3 E_-^2 ,
     	\label{eq:j_general}
     \end{equation}
     where $\eta_{1,2,3}$ are complex coefficients.
     
     Under a $\phi$ rotation,
     \begin{equation}
     	j_+' = \eta_1 e^{i2\phi}E_+^2 + \eta_2 E_+E_- + \eta_3 e^{-i2\phi}E_-^2 .
     \end{equation}
     Covariance of the vector component requires $j_+' = e^{i\phi} j_+$.
     For arbitrary $\mathbf{E}$, this can hold only if each term independently satisfies
     \begin{align}
     	\eta_1 e^{i2\phi} &= e^{i\phi} \eta_1 , \nonumber\\
     	\eta_2 &= e^{i\phi} \eta_2 , \nonumber\\
     	\eta_3 e^{-i2\phi} &= e^{i\phi} \eta_3 .
     \end{align}
     Substituting $\phi=2\pi/3$ gives
     \begin{equation}
     	\eta_1 = 0, \qquad \eta_2 = 0, \qquad \eta_3 \ \text{arbitrary}.
     \end{equation}
     Therefore, the only non-vanishing invariant structure under $C_3$ rotation is
     \begin{equation}
     	j_+ = \eta\, E_-^2, \qquad j_- = \eta^* E_+^2 ,
     \end{equation}
     where $\eta$ is a single complex parameter $\eta=a+bi$. 
     Using
     \begin{align}
     	E_-^2 = (E_x - iE_y)^2 = (E_x^2 - E_y^2) - 2iE_xE_y ,\\
     	E_+^2 = (E_x + iE_y)^2 = (E_x^2 - E_y^2) + 2iE_xE_y ,
     \end{align}
     we have
     \begin{align}
     	j_x + i j_y = (a+bi) \big[(E_x^2 - E_y^2) - 2iE_xE_y \big]\\
     	j_x - i j_y = (a-bi) \big[(E_x^2 - E_y^2) + 2iE_xE_y \big].
     \end{align}
     Separating the real and imaginary parts yields
     \begin{align}
     	j_x = a (E_x^2 - E_y^2)+2bE_x E_y, \\
     	j_y = -2a E_x E_y+b (E_x^2 - E_y^2) .
     \end{align}
     Hence the non-zero tensor components (symmetric in $j,k$) are
     \begin{align}
     	\sigma_{xxx}=a, \qquad 
     	\sigma_{xyy}=-a, \qquad 
     	\sigma_{yxy}=\sigma_{yyx}=-a\\
     	\sigma_{yyy}=-b, \qquad 
     	\sigma_{yxx}=b, \qquad 
     	\sigma_{xxy}=\sigma_{xyx}=b,
     \end{align}
     In our system, we have $\sigma_{xxx}=0$, thus $a=0$.
 
\begin{figure*}
	\centering
	\includegraphics[width=2.0\columnwidth]{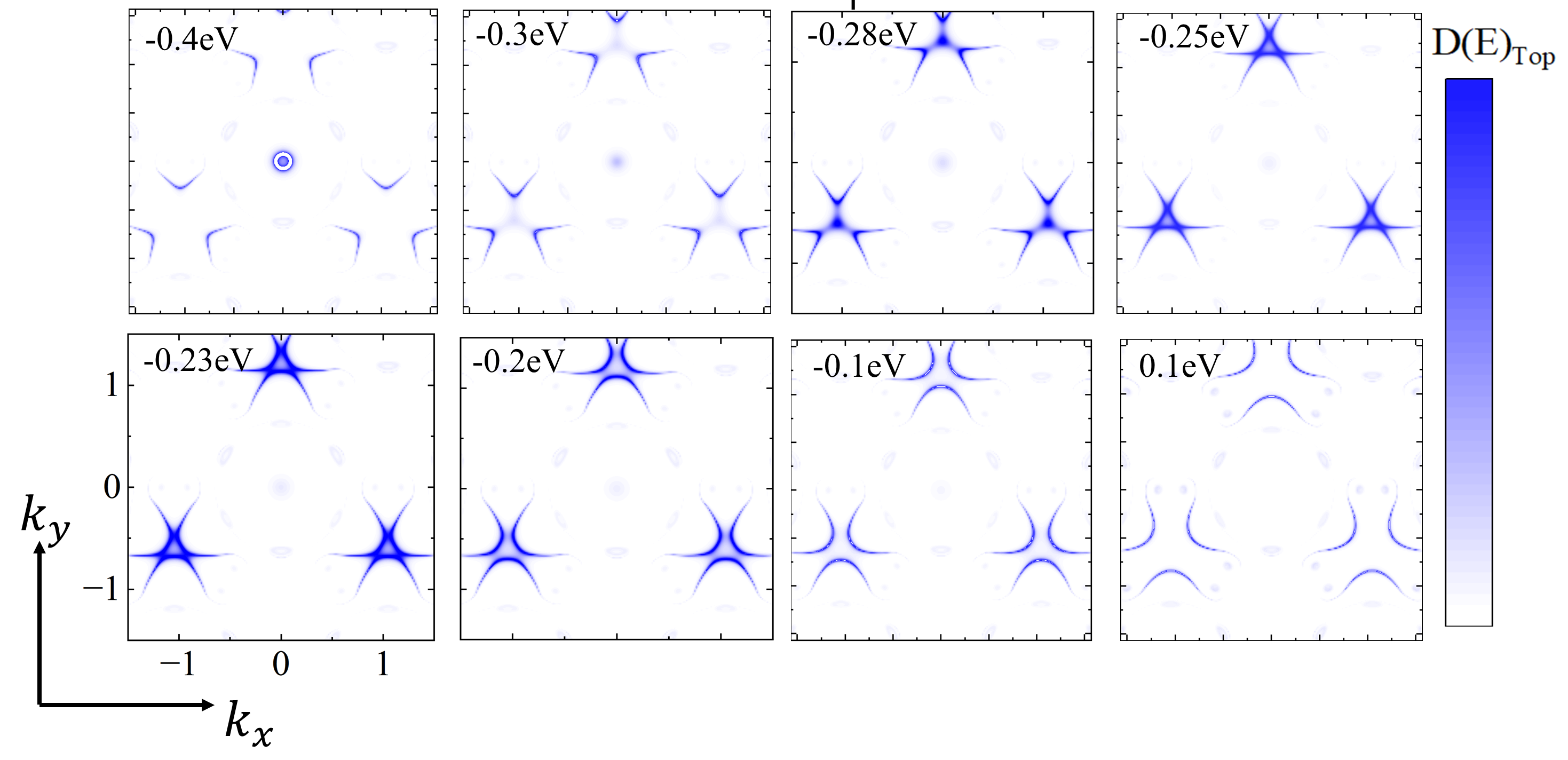}
	\caption{Evolution of top-surface state density from region I to II.
	}\label{fig7}
\end{figure*}

     For an arbitrary in-plane direction 
     $\hat{n}=(\cos\theta,\sin\theta)$ and field $\mathbf{E}=E\hat{n}$,
     \begin{align}
     	j_y &= bE^2(\cos^2\theta - \sin^2\theta), \nonumber\\
     	j_x &= 2bE^2\cos\theta\sin\theta .
     \end{align}
     The projection of $\mathbf{j}$ along $\hat{n}$ is
     \begin{align}
     	\frac{j_n}{E^2} 
     	&= \frac{\mathbf{j}\cdot\hat{n}}{E^2} 
     	= a \big[ \sin\theta(\cos^2\theta - \sin^2\theta) 
     	- 2\cos\theta\cos\theta\sin\theta \big] \nonumber\\
     	&= a(-\sin^3\theta + 3\cos^2\theta\sin\theta)
     	= a \sin(3\theta).
     \end{align}
     Therefore, the effective scalar nonlinear conductivity along $\hat{n}$ is
     \begin{equation}
     \sigma_{nnn} = a\sin(3\theta).
     \end{equation}
     
     This result shows that under $C_3$ rotational symmetry, the in-plane second-order conductivity tensor is completely determined by a single complex coefficient $\eta$, and its directional dependence follows a characteristic $\sin(3\theta)$ pattern.
    
     \section{Details of the Fermi-arc Lifshitz transition }
     In Fig.~\ref{fig7} we show details of the Fermi-arc Lifshitz transition between regions I and II in main text. At the transition energy near 
     $-0.25$eV, the two Fermi-arc configurations interchange, yielding opposite contributions to the second-order conductivity and thus a vanishing net signal. A similar scenario occurs in the II-III transition.
\end{document}